\documentclass{article}

\usepackage{arxiv}

\usepackage[utf8]{inputenc} 
\usepackage[T1]{fontenc}    
\usepackage{hyperref}       
\usepackage{url}            
\usepackage{booktabs}       
\usepackage{amsfonts}       
\usepackage{nicefrac}       
\usepackage{microtype}      
\usepackage{lipsum}
\usepackage{graphicx}
\usepackage[authoryear,round]{natbib}

\usepackage{bm}
\usepackage{dcolumn}
\usepackage{ascmac}
\usepackage{amsmath,amssymb}

\hypersetup{
    colorlinks = true,
    urlcolor   = blue,
    citecolor  = black,
}

\title{Uncertainty-Aware Spatiotemporal Super-Resolution Data Assimilation with Diffusion Models}

\author{
A. S. P. Ayapilla$^{1,\dagger,*}$, 
K. Miyashita$^{2,3,\dagger}$, 
Y. Yasuda$^{4}$, 
R. Onishi$^{1}$\\
\small
$^{1}$Supercomputing Research Center, Institute of Integrated Research,\\
Institute of Science Tokyo, Tokyo 152-8550, Japan\\
$^{2}$ADVANTEST CORPORATION, Japan\\
$^{3}$Department of Mechanical Engineering, School of Engineering,\\
Institute of Science Tokyo, Japan (during this work)\\
$^{4}$Research Institute for Value-Added-Information Generation, JAMSTEC,\\
Yokohama, Japan\\
$^{\dagger}$These authors contributed equally to this work.\\
$^{*}$Corresponding author: ayapilla.asp@scrc.iir.isct.ac.jp
}
\begin{document}
\maketitle
\begin{abstract}
Data assimilation (DA) improves prediction of chaotic systems by combining model forecasts with sparse, noisy observations. Many widely used DA methods are inherently probabilistic, requiring uncertainty quantification rather than only a single best estimate. However, accurate probabilistic DA is often computationally expensive because it relies on repeated high-resolution (HR) forecasts and large ensembles, which can limit its practical use in time-sensitive settings. In this study, a probabilistic spatiotemporal super-resolution data assimilation framework, \textbf{DiffSRDA}, is developed using denoising diffusion models and evaluated on an idealized barotropic ocean jet instability testbed. DiffSRDA is trained offline to generate short HR analysis windows conditioned on (i) a time series of low-resolution (LR) forecast frames from a numerical model and (ii) sparse HR observations. Repeated reverse diffusion sampling then produces an ensemble of HR analyses, providing both point estimates and ensemble-based uncertainty information. Despite relying only on low-cost LR forecasts, DiffSRDA achieves reconstruction quality close to that of an Ensemble Kalman Filter (EnKF) driven by HR forecasts, while improving over deterministic CNN-based SRDA baselines. The sampled ensemble also yields physically meaningful uncertainty patterns, with spread concentrated in dynamically active regions similarly to EnKF. A key practical result is that accurate base DiffSRDA cycling does not require long reverse chains, so most of the full-chain accuracy can be retained with only a small number of reverse steps, making diffusion-based SRDA computationally practical for repeated cycling. Finally, by exploiting the score-based structure of diffusion sampling, training-free observation-consistency guidance is demonstrated for deployment-time sensor-layout shifts, allowing improved use of changed observation configurations without retraining. Overall, diffusion models provide a practical, uncertainty-aware, and computationally efficient approach for spatiotemporal SRDA in chaotic fluid flows. The source code for this work is available at \url{https://github.com/Pranithhazard/DiffSRDA}.
\end{abstract}

\vspace{6pt} 

\centerline{
\begin{minipage}[t]{150mm}
\end{minipage}
}

\vspace{12pt} 

\section{Introduction}
Reliable forecasting of chaotic geophysical flows is essential in applications ranging from severe-weather preparedness to energy, aviation, and urban street-scale operations. Yet accurate prediction remains fundamentally difficult because the true flow state is only partially observed: measurements are sparse, noisy, and irregular in space and time, while forecast errors can grow rapidly in high-dimensional, strongly nonlinear dynamics. Data assimilation (DA) addresses this by combining observations with a dynamical model to estimate the evolving state and, ideally, its uncertainty. In practice, however, obtaining high-resolution analyses with reliable uncertainty quantification is often computationally expensive, especially when the forecast model must be evolved at fine resolution and within an ensemble. This creates a major bottleneck for time-sensitive applications and motivates DA strategies that can exploit cheaper low-resolution forecasts while still producing accurate high-resolution analyses.

This bottleneck is precisely where scientific machine learning becomes attractive. The growing availability of large observational datasets and high-fidelity numerical simulations makes it possible to learn useful structure offline from expensive data and then reuse it later for much cheaper online inference. In fluid dynamics, this has motivated the adaptation of ideas such as super-resolution and learned compression to reconstruct or represent high-dimensional turbulent flows from limited or low-fidelity information \citep{onishi2019super,fukami2019super,fukami2021global,yasuda2022super,yasuda2023super}. Related work has also used representation learning to build compact latent descriptions and reduced-order surrogate models for strongly nonlinear flows \citep{lusch2018deep,brunton2021modern,fukami2023grasping,solera2024beta}. The particularly useful aspect of these approaches is that they can extract multiscale structure and nonlinear statistical relationships from data that would be too expensive to exploit directly inside repeated inference loops. These ideas are also relevant to data assimilation, where the goal is not only forecasting but also efficient inference of high-resolution states from partial observations. 

\emph{Super-resolution data assimilation} (SRDA) aims to run cheaper low-resolution (LR) forecasts while still producing high-resolution (HR) analyses. \citet{barthelemy2022super} introduced SRDA by using a supervised network to super-resolve LR forecasts to HR fields, followed by an ensemble Kalman filter update with HR observations, so learning is used for super-resolution while the assimilation step remains classical. More recently, \citet{yasuda2023spatio} proposed a simultaneous SRDA operator that predicts a short HR analysis window directly from an LR forecast window and sparse HR point observations, and \citet{notsu2025four} extended this formulation to three-dimensional settings. These learned analysis operators can greatly reduce cycling cost, but many are deterministic and therefore do not directly provide uncertainty estimates, even though uncertainty is central to DA. To address this, \citet{yasuda2025unsupervised} proposed an unsupervised SRDA formulation based on conditional variational autoencoders (CVAEs), deriving an evidence lower bound (ELBO) for DA that connects to 3D-Var and leveraging the non-locality of super-resolution to induce background covariance structure from data. While this provides a clear route to probabilistic SRDA, CVAE families can still be restrictive when the conditional analysis distribution is strongly non-Gaussian, motivating diffusion and score-based generative models as a complementary approach with richer uncertainty representations.

Diffusion models, or more generally score-based generative models, provide a practical way to represent probabilistic priors for high-dimensional fields \citep{sohl2015deep,ho2020denoising,song2020score,lai2025principles}. In their common formulation, a forward process gradually perturbs clean samples with Gaussian noise, and a learned reverse process removes that noise to generate samples from the data distribution \citep{ho2020denoising}. The same idea can be written in continuous time as a reverse-time SDE driven by the score of the noisy marginals \citep{song2020score}. In the literature, diffusion models are commonly trained using the denoising formulation introduced in Denoising Diffusion Probabilistic Models (DDPM), while inference can also be carried out with deterministic or partially deterministic samplers such as Denoising Diffusion Implicit Models (DDIM) \citep{song2020denoising}. Latent diffusion approaches perform the generative process in a learned compressed space and decode back to physical space, which can reduce computational cost at high resolutions \citep{rombach2022high}. Together, these ingredients make diffusion models attractive when one needs not only a point estimate but also a controlled way to represent uncertainty through ensembles of samples.

These ideas connect directly to diffusion-based data assimilation, where score-based or diffusion models are combined with an observation model at inference to generate posterior samples \citep{rozet2023score,huang2024diffda,du2024conditional,shysheya2024conditional,wang2025phyda}. An important point is that the Bayesian posterior implicitly targeted by a diffusion-based DA procedure depends on how the model is trained and how sampling is performed, including choices that correspond to climatological priors, cycling priors, or extended-likelihood constructions that blend model information and observations \citep{hodyss2026using}. This viewpoint helps position diffusion-based SRDA as a probabilistic analysis framework that can fuse multiple information sources while remaining computationally practical.

Therefore, in this paper, we present \textbf{DiffSRDA}, a conditional diffusion-based framework for super-resolution data assimilation (SRDA) that infers a distribution over high-resolution analysis windows from a short low-resolution forecast window and sparse high-resolution observations. This enables both accurate high-resolution reconstructions and practical ensemble-based uncertainty estimates through repeated conditional sampling. As a secondary contribution, we introduce \emph{training-free observation-consistency guidance} as a lightweight deployment-time mechanism for incorporating additional or shifted observations without retraining the diffusion model. This is motivated by realistic sensor-evolution scenarios in which the observation layout or density available at deployment differs from that used during training. The proposed guidance is inspired by diffusion-posterior-sampling-style inference-time likelihood corrections \citep{chung2022diffusion}, but is specialized to the sparse on-grid masking operator arising in SRDA, making it inexpensive and easy to incorporate into reverse diffusion sampling. In addition, we evaluate a latent diffusion variant as an ablation study, in which the same conditional sampling problem is solved in a compressed latent space to reduce the diffusion-state dimension. To keep the main narrative focused on the primary DiffSRDA framework, we present this latent-space ablation in Appendix~\ref{app:SRDA_ldm}. Overall, we ask: \emph{How effectively can diffusion models perform super-resolution and data assimilation simultaneously, while also providing practical uncertainty estimates for chaotic multiscale flows?}

The remainder of the paper is organized as follows. Section~\ref{sec:Problem_Formulation} introduces the SRDA problem setting, baseline methods, and diffusion-model background, and Section~\ref{sec:SRDA_condition_diffusion} presents the proposed DiffSRDA framework and its forecast--analysis pipeline. Section~\ref{sec:training_eval_setup} describes the barotropic-jet testbed, training setup, evaluation protocol, and implementation details. Sections~\ref{sec:results} and~\ref{sec:uq} present the reconstruction results and uncertainty quantification, respectively. Section~\ref{sec:obsguid} introduces deployment-time training-free observation-consistency guidance and evaluates its effect under sensor-layout shifts. Finally, Section~\ref{sec:conclusions} concludes with a discussion and future directions.

\section{Problem formulation and methodology}
\label{sec:Problem_Formulation}

\subsection{Problem setting, previous SRDA methods, and overview of DiffSRDA}
\label{sec:Outline_and_Baseline_Methods}

\begin{figure}
  \centering
  \includegraphics[width=1.0\textwidth]{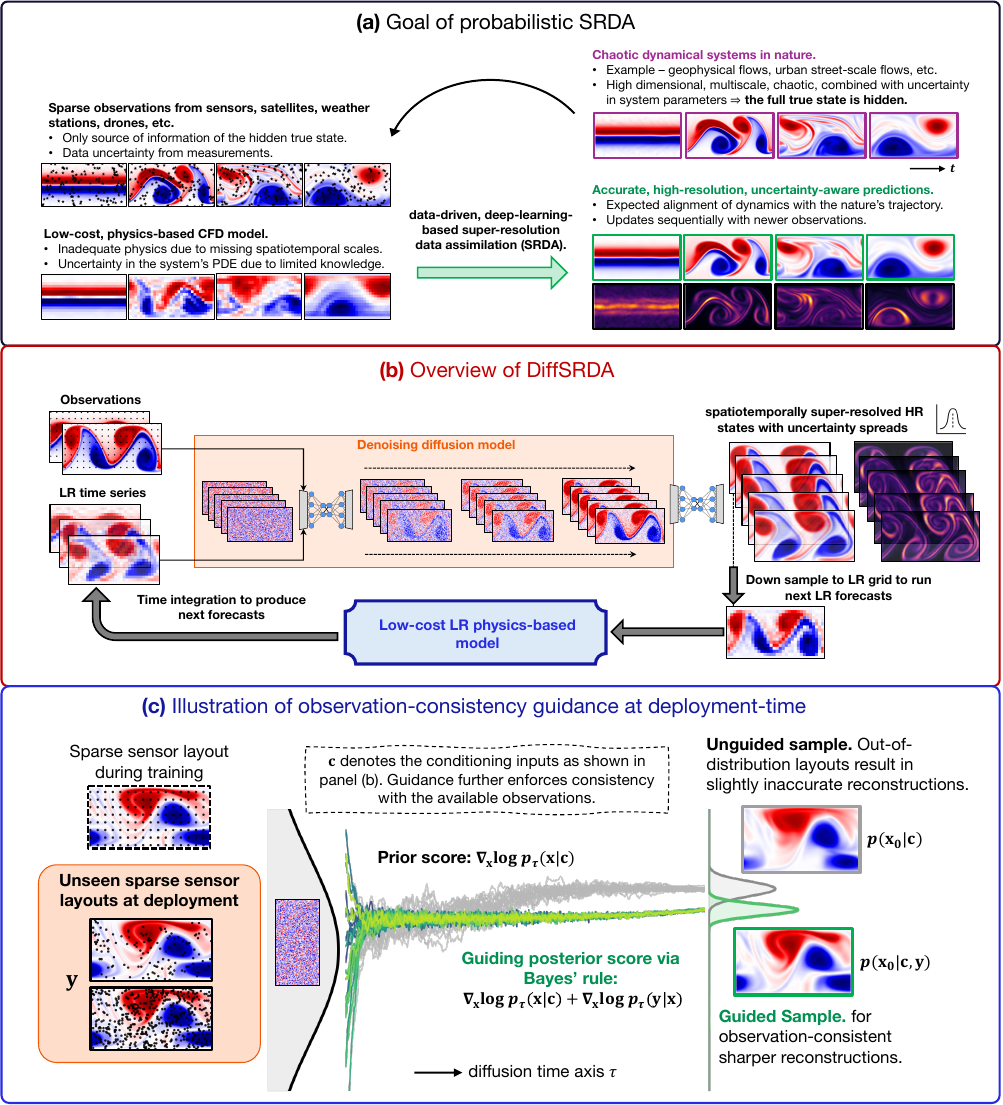}
  \caption{Conceptual overview of the probabilistic super-resolution data assimilation problem and the proposed DiffSRDA framework. \textbf{(a) Goal of probabilistic SRDA.} Sparse and noisy observations provide incomplete information about an underlying high-dimensional chaotic flow, while a low-cost physics-based low-resolution (LR) model captures only coarse dynamics. The goal is to combine both sources of information to infer accurate, high-resolution (HR), uncertainty-aware flow states. \textbf{(b) Overview of DiffSRDA.} Conditioned on sparse observations and an LR forecast time series, the denoising diffusion model generates spatiotemporally super-resolved HR state sequences together with sample-based uncertainty. The reconstructed HR states can then be downsampled and used to initialize the next LR forecast cycle. \textbf{(c) Deployment-time score-based observation-consistency guidance for unseen sensor layouts.} When the observation configuration differs from that used during training, reverse diffusion sampling is guided using the observation likelihood so that samples are nudged from the learned conditional distribution toward greater consistency with the deployment-time observations, yielding sharper and more accurate reconstructions.}
\label{fig:various_srda_principles}
\end{figure}

We begin by formulating the super-resolution data assimilation (SRDA) problem considered in this paper. Our setting follows the standard sequential forecast--analysis view of data assimilation, as in filtering methods: at each cycle, a forecast model produces a background estimate, newly available observations are assimilated to obtain an analysis, and that updated analysis is then used to initialize the next forecast cycle. In SRDA, this forecast--analysis loop is modified so that time integration is carried out with a low-resolution (LR) model for computational efficiency, while the analysis target remains a high-resolution (HR) state. The goal is therefore not only to recover missing small-scale structure, but also to keep the predicted trajectory aligned with the evolving underlying nature run as new observations arrive over time. This sequential viewpoint is especially important in chaotic multiscale systems, where transient structures, instability growth, and nonlinear interactions can quickly amplify small state errors.

Figure~\ref{fig:various_srda_principles}(a) summarizes the broader motivation. In realistic geophysical settings, the full HR state of the system is hidden. Sparse and noisy observations provide only partial information, while a low-cost physics-based model, although dynamically informative, cannot resolve all relevant spatiotemporal scales and is itself imperfect. In such settings, one cannot expect direct simulation alone to recover the true trajectory, especially under sensitivity to initial conditions, partial boundary and forcing information, model-form uncertainty, and the prohibitive cost of repeatedly evolving HR ensembles. SRDA aims to bridge this gap by combining sparse observational information with inexpensive LR dynamics to infer HR analyses that are both accurate and computationally practical.

Within this general setting, Figure~\ref{fig:various_srda_principles} also helps position the representative baseline approaches considered in this paper. One natural reference baseline is the standard ensemble Kalman filter applied fully in the high-resolution (HR) space, in which uncertainty is represented by an ensemble of HR states, each ensemble member is advanced with an HR forecast model, and the analysis update is computed from the ensemble covariance through the Kalman gain; throughout this paper, we refer to this baseline as \emph{EnKF-HR} for convenience. A second baseline follows the super-resolution ensemble Kalman filter strategy of \citet{barthelemy2022super}, in which the forecast ensemble is generated using a low-resolution (LR) model and then mapped to HR before the EnKF update through a super-resolution (SR) operator; this retains the classical EnKF structure while reducing forecast cost because only LR dynamics are evolved online, and in the present work we use bicubic interpolation for this LR-to-HR mapping rather than a pretrained neural SR model so that the baseline cleanly isolates the effect of LR forecasting combined with an HR EnKF analysis step; for convenience, we denote this baseline by \emph{EnKF-SR}. A more direct learned alternative is super-resolution data assimilation (SRDA) \citep{yasuda2023spatio}, in which the analysis operator itself is replaced by a trained neural network that takes LR forecast information together with sparse observations and directly outputs an HR analysis state, or a short HR analysis window, in a single forward pass. This can substantially reduce the cost of cycling, but in its basic form the learned SRDA operator is deterministic and therefore does not directly provide uncertainty information.

That limitation motivates the probabilistic SRDA framework developed here. In chaotic and multiscale flows, the same coarse forecast information and sparse observations may remain consistent with multiple plausible HR realizations, especially once instabilities develop and nonlinear structures emerge. A useful SRDA method should therefore provide not only an HR reconstruction, but also a practical representation of uncertainty over possible HR analyses. In this work, we address this need by replacing the deterministic SRDA operator with a conditional diffusion model, yielding \textbf{DiffSRDA}, which defines a distribution over HR analysis windows conditioned on LR forecast information and observations.

Figure~\ref{fig:various_srda_principles}(b) provides a high-level overview of DiffSRDA. At each assimilation cycle, a short LR forecast window together with sparse observations is used as conditioning information, and the denoising diffusion model generates a spatiotemporally super-resolved HR analysis window. Repeated conditional sampling then provides both a point estimate, for example through the sample mean, and an ensemble-based estimate of predictive uncertainty. The resulting HR analysis can be downsampled and fed back to the LR model to initialize the next forecast cycle, thus preserving the sequential filtering structure of SRDA. 

Figure~\ref{fig:various_srda_principles}(c) illustrates an additional practical feature of the diffusion-based formulation. Because reverse diffusion sampling is score-based, it can be modified at deployment time using an observation-consistency guidance term. This provides a lightweight mechanism for incorporating additional or shifted observations when the sensor configuration available at deployment differs from that used during training, without retraining the diffusion model itself. We only give this intuitive overview here; the full DiffSRDA formulation is presented in Section~\ref{sec:SRDA_condition_diffusion}, and the guidance method and its derivation are developed later in Section~\ref{sec:obsguid} and Appendix~\ref{app:likelihood_guidance_derivation}.

\subsection{Denoising diffusion models}
\label{sec:Denoising_Diffusion_Models}

\paragraph{Score-based view of diffusion models:}
We briefly summarize the conditional denoising diffusion model used in this study and introduce the notation needed for later derivations. Let $\mathbf{X}_0$ denote the target HR field. In our setting, $\mathbf{X}_0$ represents a short HR window of vorticity frames (described in more detail in the next section). A diffusion model defines (i) a forward noising process that progressively perturbs $\mathbf{X}_0$ into a noise-like variable $\mathbf{X}_T$, and (ii) a learned reverse-time process that maps noise back into an HR sample consistent with the conditioning information.

It is helpful to briefly connect discrete diffusion models to continuous-time diffusion theory. To avoid introducing new symbols, we use the same diffusion-time notation $\tau$ as in the discrete formulation and interpret it here as a continuous variable $\tau\in[0,1]$. A forward diffusion can be written as an It\^o stochastic differential equation (SDE) driven by a Wiener process $W_\tau$,
\begin{equation}
\mathrm{d}\mathbf{X}_\tau = f(\mathbf{X}_\tau,\tau)\,\mathrm{d}\tau + G(\tau)\,\mathrm{d}W_\tau,
\label{eq:sde_forward_short}
\end{equation}
whose marginals satisfy a Fokker--Planck equation. A key result is that the time-reversed process is again a diffusion, with an additional drift term that depends on the score of the conditional marginals \citep{anderson1982reverse,song2020score}:
\begin{equation}
\mathrm{d}\mathbf{X}_\tau
=
\Bigl(
f(\mathbf{X}_\tau,\tau) - D(\tau)\,\nabla_{\mathbf{X}_\tau} \log p_\tau(\mathbf{X}_\tau\mid \mathbf{c})
\Bigr)\mathrm{d}\tau
+
G(\tau)\,\mathrm{d}\bar W_\tau,
\qquad
D(\tau)=G(\tau)G(\tau)^\top,
\label{eq:sde_reverse_short}
\end{equation}
where $\bar W_\tau$ is a Wiener process when integrating backward in $\tau$. This score term motivates learning $\nabla_{\mathbf{X}_\tau}\log p_\tau(\mathbf{X}_\tau\mid \mathbf{c})$ from data, and the discrete Gaussian Markov chain used below can be viewed as a particular time discretization of these dynamics.

Coming to the discrete picture, we index diffusion time by $\tau\in\{1,\dots,T\}$, where $T$ is the total number of diffusion steps used in the forward process. The forward process is a Markov chain that adds Gaussian noise,
\begin{equation}
q(\mathbf{X}_\tau \mid \mathbf{X}_{\tau-1})
=
\mathcal{N}\!\left(\sqrt{1-\beta_\tau}\,\mathbf{X}_{\tau-1},\, \beta_\tau I\right),
\label{eq:forward_diffusion}
\end{equation}
with a fixed variance schedule $\{\beta_\tau\}_{\tau=1}^T$. Define $\alpha_\tau = 1-\beta_\tau$ and $\bar\alpha_\tau = \prod_{s=1}^{\tau}\alpha_s$. Then $\mathbf{X}_\tau$ can be sampled directly from $\mathbf{X}_0$ as
\begin{equation}
\mathbf{X}_\tau
=
\sqrt{\bar\alpha_\tau}\,\mathbf{X}_0
+
\sqrt{1-\bar\alpha_\tau}\,\boldsymbol{\varepsilon},
\qquad
\boldsymbol{\varepsilon}\sim\mathcal{N}(0,I),
\label{eq:direct_noising}
\end{equation}
which is the form used during training.

\paragraph{Conditional denoising objective:}
SRDA requires a conditional generative model. We denote the conditioning information by $\mathbf{c}$, which collects the LR forecast features and the processed observation channels introduced in Sections~\ref{sec:Outline_and_Baseline_Methods} and~\ref{sec:SRDA_condition_diffusion}. The denoiser is a neural network $\boldsymbol{\varepsilon}_\theta(\cdot)$ that takes $(\mathbf{X}_\tau,\tau,\mathbf{c})$ and predicts the noise used to generate $\mathbf{X}_\tau$. The standard DDPM training objective minimizes the mean-squared error between the true noise and the predicted noise,
\begin{equation}
\mathcal{L}(\theta)
=
\mathbb{E}_{\tau,\,\mathbf{X}_0,\,\boldsymbol{\varepsilon}}
\left[
\left\|
\boldsymbol{\varepsilon}-\boldsymbol{\varepsilon}_{\theta}(\mathbf{X}_{\tau},\tau,\mathbf{c})
\right\|_2^2
\right],
\label{eq:ddm_loss}
\end{equation}
where $\tau$ is typically sampled uniformly from $\{1,\dots,T\}$ and $\boldsymbol{\varepsilon}\sim\mathcal{N}(0,I)$. In our implementation, however, the denoiser is trained with an $L^1$ noise-prediction loss rather than this canonical $L^2$ form.

\paragraph{Reverse-diffusion sampling:}
At inference, we draw $\mathbf{X}_T\sim\mathcal{N}(0,I)$ and run a reverse-time procedure conditioned on $\mathbf{c}$ to obtain an HR analysis sample. 
The forward noising relation
\begin{equation}
\mathbf{X}_\tau = \sqrt{\bar\alpha_\tau}\,\mathbf{X}_0 + \sqrt{1-\bar\alpha_\tau}\,\boldsymbol{\varepsilon},
\qquad \boldsymbol{\varepsilon}\sim\mathcal{N}(0,I),
\label{eq:forward_relation}
\end{equation}
implies a simple connection between the denoiser output and the \emph{score} of the conditional noisy marginal. We define the conditional score at diffusion time $\tau$ as
\begin{equation}
s_\tau(\mathbf{X}_\tau,\mathbf{c}) \;\triangleq\; \nabla_{\mathbf{X}_\tau}\log p_\tau(\mathbf{X}_\tau\mid \mathbf{c}),
\label{eq:score_def}
\end{equation}
where $p_\tau(\cdot\mid \mathbf{c})$ denotes the marginal density of $\mathbf{X}_\tau$ induced by equation \eqref{eq:forward_relation} under the data distribution and conditioning inputs $\mathbf{c}$. For the parameterization in equation \eqref{eq:forward_relation}, the standard $\boldsymbol{\varepsilon}$-prediction network provides a direct estimator of the score:
\begin{equation}
s_\tau(\mathbf{X}_\tau,\mathbf{c})\;\approx\; -\,\frac{1}{\sqrt{1-\bar\alpha_\tau}}\;\boldsymbol{\varepsilon}_\theta(\mathbf{X}_\tau,\tau,\mathbf{c}).
\label{eq:score_from_eps}
\end{equation}
Using $\boldsymbol{\varepsilon}_\theta$, we form the usual clean-sample estimate
\begin{equation}
\hat{\mathbf{X}}_0(\mathbf{X}_\tau)
=
\frac{1}{\sqrt{\bar\alpha_\tau}}
\left(
\mathbf{X}_\tau - \sqrt{1-\bar\alpha_\tau}\,\boldsymbol{\varepsilon}_\theta(\mathbf{X}_\tau,\tau,\mathbf{c})
\right).
\label{eq:x0_hat}
\end{equation}
Combining equation \eqref{eq:x0_hat} with equation \eqref{eq:score_from_eps} yields the equivalent score form
\begin{equation}
\hat{\mathbf{X}}_0(\mathbf{X}_\tau)
=
\frac{1}{\sqrt{\bar\alpha_\tau}}
\left(
\mathbf{X}_\tau + (1-\bar\alpha_\tau)\, s_\tau(\mathbf{X}_\tau,\mathbf{c})
\right),
\label{eq:x0_hat_score_form}
\end{equation}
which is a Tweedie-type relation: the posterior mean of the clean variable is obtained by shifting the noisy sample in the direction of the score, scaled by the noise variance.

Reverse-time sampling can then be written directly in terms of $\hat{\mathbf{X}}_0(\mathbf{X}_\tau)$ and $\boldsymbol{\varepsilon}_\theta$. We use the unified update
\begin{equation}
\mathbf{X}_{\tau-1}
=
\sqrt{\bar\alpha_{\tau-1}}\,\hat{\mathbf{X}}_0(\mathbf{X}_\tau)
+ \sqrt{1-\bar\alpha_{\tau-1}-\sigma_\tau^2}\,\boldsymbol{\varepsilon}_\theta(\mathbf{X}_\tau,\tau,\mathbf{c})
+ \sigma_\tau \mathbf{z},
\qquad
\mathbf{z}\sim\mathcal{N}(0,I),
\label{eq:ddim_update}
\end{equation}
with
\begin{equation}
\sigma_\tau
=
\eta\,
\sqrt{\frac{1-\bar\alpha_{\tau-1}}{1-\bar\alpha_{\tau}}}\,
\sqrt{1-\frac{\bar\alpha_{\tau}}{\bar\alpha_{\tau-1}}}\,,
\qquad \eta\in[0,1].
\label{eq:eta_sigma}
\end{equation}
When $\eta=0$, the update is deterministic (DDIM) given the initial draw $\mathbf{X}_T$, while $\eta=1$ recovers a stochastic sampler consistent with the forward variance schedule (DDPM-style sampling).

To reduce inference cost, we can apply timestep respacing. Instead of traversing the full reverse chain, we select a reduced set of diffusion times
\[
T=\tau_K > \tau_{K-1} > \cdots > \tau_1 > \tau_0=0,
\]
construct a shortened reverse process whose cumulative noise levels match the corresponding values $(\bar\alpha_{\tau_K},\bar\alpha_{\tau_{K-1}},\ldots,\bar\alpha_{\tau_0})$, and then apply equation~\eqref{eq:ddim_update} on this reduced grid using consecutive pairs $(\tau_k,\tau_{k-1})$. We denote the number of reverse updates by $K$ and refer to it as the number of reverse sampling steps.

\subsection{Sequential probabilistic forecasting using DiffSRDA}
\label{sec:SRDA_condition_diffusion}

\begin{figure}
  \centering
  \includegraphics[width=1.0\textwidth]{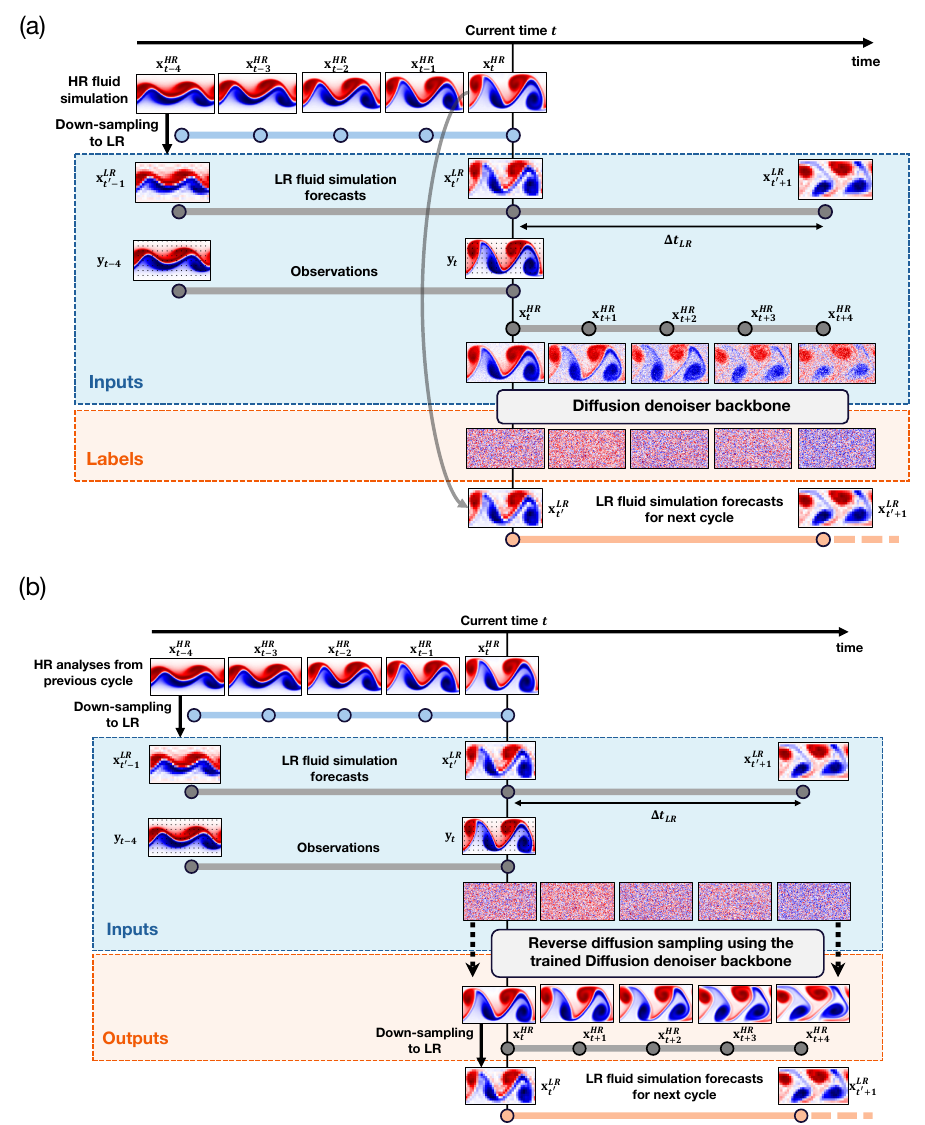}
  \caption{Schematic of one assimilation cycle in (a) training DiffSRDA and (b) inference with the trained DiffSRDA model. HR states are denoted by $\mathbf{x}^{\mathrm{HR}}_t$ at HR output time index $t$, LR forecast states by $\mathbf{x}^{\mathrm{LR}}_{t'}$ at LR forecast-output index $t'$, and observations by $\mathbf{y}_t$ at HR output time index $t$. The HR and LR states used in the SRDA pipeline are aligned at common output times, while the LR and HR solvers use different internal time steps.}
\label{fig:SRDA_training_inference}
\end{figure}

We now describe the full methodological pipeline of DiffSRDA, i.e., how the pixel-space conditional diffusion model is used as an SRDA operator within a standard forecast--analysis cycle. We first summarize the offline training setup, in which the model is trained on HR analysis windows conditioned on low-cost forecast information and sparse observations, and then describe the corresponding inference-time cycle, in which the trained model is sampled to generate HR analyses during evaluation. Figure~\ref{fig:SRDA_training_inference} provides a schematic overview of these two procedures.

We denote HR states by $\mathbf{x}^{\mathrm{HR}}_t$ at HR output time index $t$, and LR forecast states by $\mathbf{x}^{\mathrm{LR}}_{t'}$ at LR forecast-output index $t'$. The HR and LR states used in the SRDA pipeline are aligned at common output times, while the LR and HR solvers use different internal time steps. In our setup, one assimilation interval spans four HR output steps.

Figure~\ref{fig:SRDA_training_inference}(a) illustrates one assimilation cycle during training. At HR time $t$, the conditioning inputs consist of (i) a short LR forecast sequence centered around the aligned LR time index $t'$ and (ii) sparse HR observations at the assimilation times. The supervised target is a length-five HR analysis window,
\begin{equation}
\mathbf{X} \;\equiv\; \bigl[\mathbf{x}_t^{\mathrm{HR}},\mathbf{x}_{t+1}^{\mathrm{HR}},\ldots,\mathbf{x}_{t+4}^{\mathrm{HR}}\bigr],
\label{eq:srda_window_def_cycle}
\end{equation}
represented as five channels that are denoised jointly by a single conditional diffusion model. In this way, the model learns a conditional distribution over HR windows given the low-cost forecast information and the observations.

Figure~\ref{fig:SRDA_training_inference}(b) shows the corresponding inference-time cycle with the trained model. At the current HR time $t$, the same types of conditioning inputs are provided, and reverse diffusion sampling starts from Gaussian noise to generate an HR analysis window $\hat{\mathbf{X}}$. Once sampling is complete, the current-time HR analysis $\hat{\mathbf{x}}^{\mathrm{HR}}_t$ is downsampled to the LR grid and used to initialize the next LR forecast, which then provides the background information for the subsequent assimilation cycle.

These schematics motivate the probabilistic formulation used throughout the paper. The analysis variable is the HR window $\mathbf{X}$, the conditioning information is the pair $(\mathbf{f},\mathbf{y})$ consisting of the LR forecast sequence $\mathbf{f}$ and the HR observations $\mathbf{y}$, and DiffSRDA defines a conditional generator $p_\theta(\mathbf{X}\mid \mathbf{f},\mathbf{y})$.

It is also important to note that we adopt a four-dimensional SRDA (4D-SRDA) design, in which the inference model consumes a short time series of LR model states together with observations and outputs an HR time series, while only the current HR state is downsampled and fed back to initialize the next forecast cycle. This windowed formulation provides temporal context that makes the HR inference problem better posed than single-frame super-resolution, promotes short-range temporal coherence, and reduces ambiguity in the recovered small-scale structure of chaotic flows. From a probabilistic perspective, the diffusion model therefore learns an amortized conditional score over a fixed-lag HR segment (here, five frames) given the LR-forecast/observation context, while still operating within a sequential cycling framework rather than a full-trajectory smoothing formulation.

\section{Barotropic-jet testbed, training setup, and evaluation}
\label{sec:training_eval_setup}

This section describes the dynamical-system testbed, numerical configurations, training setup, and evaluation protocol used in the experiments. Since the forecast--analysis cycle and the DiffSRDA formulation were introduced in Section~\ref{sec:Problem_Formulation} and Section~\ref{sec:SRDA_condition_diffusion}, we focus here on the remaining ingredients needed to reproduce the results, namely the barotropic-jet model, its multi-resolution dataset construction, the observation setup, and the training and sampling configurations. The U-Net denoiser backbone architectures used in this study are summarized in Appendix~\ref{app:architectures}.

\subsection{Barotropic-jet testbed}
\label{sec:Testbed_Barotropic_Jet_Instability}

Following \citet{ribstein2014barotropic,david2017statistical}, we use an idealized barotropic ocean jet as a testbed for super-resolution data assimilation in a chaotic multiscale geophysical-flow setting. The system evolves on a domain that is periodic in $x$ and bounded in $y$, and it exhibits barotropic instability, vortex roll-up, nonlinear filamentation, vortex merging, and a late-time statistically stationary regime dominated by a few long-lived westward-propagating coherent vortices. In our experiments, the ultra-high-resolution (UHR) simulation provides the reference ``nature'' trajectory, while lower-resolution simulations are used to define the low-cost forecast model and the target analysis grid (see section \ref{sec:data_grids_obs} below). Figure~\ref{fig:barotrpicjet_sim} illustrates several UHR realizations generated from different perturbation initial conditions. Although these realizations share the same broad sequence of instability growth and nonlinear evolution, they diverge markedly once the instability develops, highlighting the chaotic sensitivity of the testbed.

\subsubsection{Governing equations and instability mechanism}

The model is written in terms of the vorticity $\omega(x,y,t)$ and streamfunction $\psi(x,y,t)$:
\begin{align}
  \frac{\partial \omega}{\partial t}
  + u \frac{\partial \omega}{\partial x}
  + v \frac{\partial \omega}{\partial y}
  + f_y\, v
  &= - r\,\omega
     - \nu \Delta^2 \omega
     - \frac{d \Xi(y)}{dy},
  \label{eq:governing_equation}
\end{align}
\begin{align}
  \Delta \psi = \omega,
  \label{eq:streamline_function}
\end{align}
\begin{align}
  (u, v)
  = \left(
        -\frac{\partial \psi}{\partial y},
        \frac{\partial \psi}{\partial x}
    \right),
  \label{eq:streamlines_and_velocities}
\end{align}
where the left-hand side of equation \eqref{eq:governing_equation} represents advection of vorticity together with a constant background vorticity gradient $f_y$ (the beta-plane term, written here as $f_y$ to avoid conflict with the diffusion-model noise schedule). The right-hand side contains linear drag with coefficient $r$, hyperviscosity with coefficient $\nu$, and a prescribed meridional forcing through the derivative of $\Xi(y)$.

The initial jet is specified through a zonal velocity profile $U(y)$, and the forcing $\Xi(y)$ is chosen with the same functional form:
\begin{align}
    U(y)
    =
    U_0
    \left[
        \mathrm{sech}^2\!\left(\frac{y - y_0}{\delta}\right)
        - C
    \right],
    \label{eq:u_y}
\end{align}
and
\begin{align}
    \Xi(y)
    =
    \Xi_0
    \left[
        \mathrm{sech}^2\!\left(\frac{y - y_0}{\delta}\right)
        - C
    \right],
    \label{eq:Xi_y}
\end{align}
where $y_0$ and $\delta$ control the jet center and width, $U_0$ and $\Xi_0$ set the amplitudes, and the constant $C$ is chosen so that the $y$-integrals of $U(y)$ and $\Xi(y)$ vanish.

\begin{figure}
  \centering
  \includegraphics[width=1.0\textwidth]{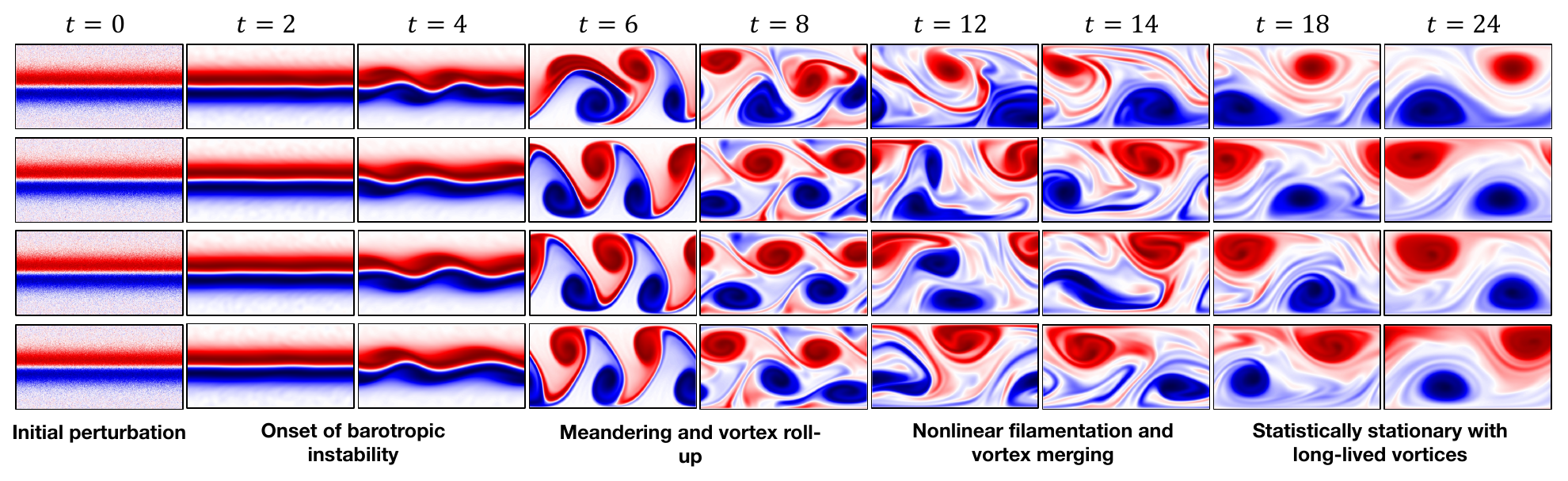}
  \caption{Evolution of the barotropic jet instability in the ultra-high-resolution (UHR) system for multiple realizations with different perturbation initial conditions. Each row shows vorticity snapshots from a distinct trajectory at selected times $t \in [0,24]$. Although all realizations exhibit the same broad sequence of barotropic-instability dynamics---initially weak perturbation growth, onset of jet meandering, roll-up into coherent vortices, nonlinear filamentation and secondary structure formation, vortex interaction and merging, and finally a late-time statistically stationary regime dominated by a few long-lived westward-propagating vortices---the detailed flow fields and exact trajectories diverge substantially across realizations once the instability develops. This sensitivity highlights the chaotic nature of the system and motivates probabilistic trajectory prediction, uncertainty quantification, and data assimilation for alignment with the underlying nature trajectory.}
\label{fig:barotrpicjet_sim}
\end{figure}

For a barotropic jet on a beta plane, a necessary condition for instability is the Rayleigh--Kuo criterion, namely that the meridional gradient of absolute vorticity,
\[
f_y - \frac{d^2U}{dy^2},
\]
changes sign somewhere in the domain. With the parameter values used here, this condition is satisfied, so the base jet is barotropically unstable. Small perturbations therefore grow into meanders, roll up into coherent vortices, generate thin filaments and smaller-scale structures through nonlinear interactions, and eventually enter a statistically stationary vortex-dominated regime. This common qualitative progression is visible across all trajectories in Figure~\ref{fig:barotrpicjet_sim}, even though the detailed flow fields and exact trajectories differ substantially from one realization to another after instability onset.

\subsubsection{Numerical method and parameter settings}

We solve equations \eqref{eq:governing_equation}--\eqref{eq:streamlines_and_velocities} on the domain $0 \le x \le 2\pi$ and $0 \le y \le \pi$ using a pseudo-spectral method. Periodic boundary conditions are imposed in the $x$ direction, while fixed-wall boundary conditions are imposed in the $y$ direction, with the corresponding streamfunction satisfying $\psi(x,0,t)=\psi(x,\pi,t)=0$. In practice, fields are represented with Fourier modes in $x$ and sine modes in $y$, truncated at cutoff wavenumbers $(K,L)$ depending on the grid resolution. Nonlinear advection terms are evaluated in physical space, and transforms between spectral and physical representations are computed efficiently using fast Fourier transforms. Time integration is performed with a modified Euler scheme.

The initial vorticity field is generated by superposing small random perturbations onto the prescribed zonal jet. Following \citet{david2017statistical}, these perturbations are introduced in spectral space using random phases and amplitudes drawn from Gaussian distributions with standard deviations $\sigma_{\mathrm{phase}}$ and $\sigma_{\mathrm{amplitude}}$, respectively. These perturbations are small enough that the early evolution is initially close to the base jet, but they trigger different nonlinear trajectories once the barotropic instability amplifies.

\begin{table}
    \caption{Parameters of the governing equations.}
    \label{table:params_Cd}
    \centering
    \resizebox{\textwidth}{!}{%
    \begin{tabular}{lccccccccccc}
        \hline
        Parameter 
        & Domain in $x$
        & Domain in $y$
        & $f_y$
        & $r$
        & $\nu$
        & $\Xi_0$
        & $y_0$
        & $\delta$
        & $U_0$
        & $\sigma_{\rm phase}$
        & $\sigma_{\rm amplitude}$ \\
        \hline
        Value
        & $[0,2\pi]$
        & $[0,\pi]$
        & $0.1$
        & $1\times 10^{-2}$
        & $1\times 10^{-5}$
        & $0.3$
        & $\pi/2$
        & $0.4$
        & $3.0$
        & $\pi$
        & $2.5\times 10^{-3}$ \\
        \hline
    \end{tabular}%
    }
\end{table}

\subsubsection{Resolution levels, dataset construction, and observations}
\label{sec:data_grids_obs}

We run the barotropic-jet model at three spatial resolutions: low resolution (LR), high resolution (HR), and ultra-high resolution (UHR). The UHR simulation provides the reference trajectory used for verification, HR is the target grid on which DiffSRDA produces analyses, and LR is used to generate low-cost forecasts. Table~\ref{table:params_simulation} summarizes the grid sizes, spectral cutoffs, time-stepping settings, and saved output intervals at the three resolutions. Over the evaluation window $t\in[0,24]$, the initially perturbed jet evolves through meandering, vortex roll-up, filamentation, and vortex merging before reaching the late-time vortex-dominated regime shown in Figure~\ref{fig:barotrpicjet_sim}.

\begin{table}
    \centering
    \caption{Simulation settings for the barotropic jet calculations at the three spatial resolutions used in this study.}
    \label{table:params_simulation}
    \begin{tabular}{lcccc}
        \hline
        Resolution &
        \begin{tabular}{c}Grid size\\$(N_x \times N_y)$\end{tabular} &
        \begin{tabular}{c}Spectral cutoff\\$(K,L)$\end{tabular} &
        \begin{tabular}{c}Integrator step\\$\Delta t$\end{tabular} &
        \begin{tabular}{c}Saved output\\interval\end{tabular}
        \\ \hline
        LR  & $32 \times 16$     & $(10,\,10)$   & $5 \times 10^{-4}$                 & $1.00\ (= \Delta t_{\mathrm{LR,out}})$ \\
        HR  & $128 \times 64$    & $(42,\,42)$   & $1.25 \times 10^{-4}$              & $0.25\ (= \Delta t_{\mathrm{HR,out}})$ \\
        UHR & $1024 \times 512$  & $(341,\,341)$ & $1.5625 \times 10^{-5}$            & $0.25\ (= \Delta t_{\mathrm{HR,out}})$ \\
        \hline
    \end{tabular}
\end{table}

In the data-assimilation setting, we denote the discretized vorticity field on the HR grid by $\mathbf{x}$, while physical space is still described by the Cartesian coordinates $(x,y)$. Two time indices are used and kept distinct throughout the paper. Physical time on the HR grid is indexed by $t$ and sampled at $\Delta t_{\mathrm{HR,out}}=0.25$. The LR forecast model advances with the coarser output interval $\Delta t_{\mathrm{LR,out}}=1.00$ and is indexed by $t'$, so one assimilation interval spans four HR steps. Diffusion time, denoted by $\tau$, is an internal index used only for diffusion-model training and reverse-time sampling.

At each assimilation time, sparse observations are generated on the HR grid from the reference state according to
\begin{equation}
\mathbf{y} = H\mathbf{x} + \boldsymbol{\eta},
\end{equation}
where $H$ is a sparse sampling operator and $\boldsymbol{\eta}$ is additive observation noise. In practice, $H$ is implemented as a binary mask that selects the observed pixels, and noise is added only at those locations. We parameterize the observation density by the grid interval \texttt{ogiX}: for example, \texttt{ogi8} observes every $g=8$ grid points in both directions, and similarly for \texttt{ogi4}, \texttt{ogi2}, and related cases. Unless stated otherwise, the regular observation lattice is shifted by a phase offset that is fixed in time and controlled by a pseudo-random seed.

Unless stated otherwise, for observations, we use independently and identically sampled Gaussian noise with standard deviation $\sigma_{\mathrm{obs}}=0.1$ in vorticity units. Observations are available only at assimilation times, that is, every four HR output steps in the present setup. Between assimilation times, the observation field is treated as missing.

\subsection{Training setup}
\label{sec:training_config}

\paragraph{Data and normalization:}
Training uses a library of HR simulations as supervised targets for SRDA, following the same general protocol as \citet{yasuda2023spatio}. All HR training samples are normalized to $[-1,1]$ using fixed global bounds $(\omega_{\min},\omega_{\max})$ computed from the HR training data. We use $120{,}000$ training samples with batch size $64$. Validation uses batch size $250$, with $30{,}000$ samples in the pixel-space setting and $10{,}000$ samples in the latent setting. Early stopping is enforced based on the validation loss.

\paragraph{Forward diffusion schedule and objective:}
We use a linear variance schedule with $T=1000$ diffusion steps,
\begin{equation}
\beta_\tau \in \left[10^{-4},\,2\times 10^{-2}\right],
\end{equation}
and train the denoiser with an $L^1$ noise-prediction loss. Optimization uses Adam with learning rate $10^{-4}$ for up to $2000$ epochs, with early stopping using patience $50$. For validation and checkpoint selection, we use a reconstruction-style criterion obtained by running the current sampler on validation inputs and comparing the reconstructed HR output against the target. The checkpoint with the lowest validation loss is then used for evaluation.

\paragraph{SR-mixup augmentation:}
To improve robustness to distribution shifts in the LR forecast inputs, for example due to resolution- or model-mismatch effects, we apply SR-mixup \citep{yasuda2023spatio} as an LR-side data augmentation. This mitigates the tendency of learned SRDA models to overfit to the exact LR training distribution and then degrade when the deployment LR statistics drift. For each training sample, we choose a second LR forecast window that is nearby in Euclidean distance, selected from $200$ randomly drawn candidates, sample $\lambda\sim\mathrm{Beta}(2,5)$, and form the mixed LR input
\begin{equation}
\tilde{\mathbf{f}}=\lambda \mathbf{f}^{(i)}+(1-\lambda)\mathbf{f}^{(j)}.
\end{equation}
Training then uses $(\tilde{\mathbf{f}},\mathbf{y}^{(i)})$ together with the original HR target window $\mathbf{X}^{(i)}$, which encourages the model to remain stable under small but physically plausible variations of the LR conditioning. SR-mixup is applied only to the LR forecast channels and not to the observation channels.

\subsection{Sampling and evaluation}
\label{sec:sampling_eval}

At inference, samples are generated by running the reverse diffusion process conditioned on the LR forecast inputs and processed observations. During evaluation, UHR simulations are used as the reference trajectories.

For diffusion-based SRDA, an ensemble of $B$ conditional samples is obtained by repeating the reverse sampler with different initial noise realizations. Unless stated otherwise, we use $B=30$ diffusion samples. For EnKF-HR, we use $N=100$ ensemble members unless noted otherwise.

In the cycling experiments, the LR forecast is restarted from the first sampled current-time HR analysis produced by DiffSRDA at each assimilation time, for computational efficiency. Point-estimate metrics, however, are reported using the posterior mean over the $B$ conditional samples. Accordingly, the reported posterior-mean errors do not correspond to a fully posterior-mean-cycled system.

\subsection{Error metrics}
\label{sec:error_metrics}

All accuracy diagnostics are computed against the UHR reference after mapping the HR inference to the UHR grid by bicubic interpolation. Let $\mathbf{x}^{\mathrm{UHR}}_t$ denote the reference UHR state at physical time $t$, and let $\hat{\mathbf{x}}^{\mathrm{UHR}}_t$ denote the corresponding prediction mapped from HR to the UHR grid.

We report the mean absolute error ratio (MAER),
\begin{equation}
\mathrm{MAER}(t)=
\frac{\mathbb{E}_{i}\!\left[\left|\hat{\mathbf{x}}^{\mathrm{UHR}}_{t,i}-\mathbf{x}^{\mathrm{UHR}}_{t,i}\right|\right]}
{\mathbb{E}_{i}\!\left[\left|\mathbf{x}^{\mathrm{UHR}}_{t,i}\right|\right]},
\end{equation}
where $\mathbb{E}_{i}[\cdot]$ denotes averaging over UHR grid points.

To assess structural similarity, we report the multi-scale structural similarity (MSSIM) loss,
\begin{equation}
\mathrm{MSSIMLoss}(t)=1-\mathrm{MSSIM}\!\left(\hat{\mathbf{x}}^{\mathrm{UHR}}_t,\,\mathbf{x}^{\mathrm{UHR}}_t\right),
\end{equation}
using the SSIM implementation of \citet{yasuda2023spatio} with window size $11$ and a Gaussian window with $\sigma=1.5$. SSIM is computed after applying the same fixed affine rescaling to both fields (bias $=-14.5$, scale $=29.0$).

To emphasize small-scale differences, we also compute the root-mean-square error (RMSE) of the Laplacian of vorticity,
\begin{equation}
\mathrm{RMSE}\bigl(\Delta \mathbf{x}\bigr)(t)=
\sqrt{
\mathbb{E}_{i}\!\left[
\left(
\left[\Delta_h \hat{\mathbf{x}}^{\mathrm{UHR}}_t\right]_i
-
\left[\Delta_h \mathbf{x}^{\mathrm{UHR}}_t\right]_i
\right)^2
\right]
},
\end{equation}
where $\Delta_h$ denotes the discrete Laplacian on the UHR grid, with boundary treatment consistent with the numerical discretization of the barotropic jet solver, that is, periodic in $x$ and wall-consistent in $y$.

\subsection{Additional implementation details}
\label{sec:impl_baselines}

EnKF-HR and EnKF-SR use the perturbed-observation EnKF update and the same assimilation interval as DiffSRDA. EnKF-SR advances an ensemble using the LR forecast model, super-resolves each LR ensemble member to HR by bicubic interpolation, and applies the EnKF update in HR space, following \citet{barthelemy2022super}. For the baseline observation setting, observation perturbations in the EnKF update use standard deviation $\sigma_{\mathrm{obs}}=0.10$, and covariance localization uses the Gaspari--Cohn function with localization radii $0.5$ in both $x$ and $y$ in nondimensional coordinates. We use a multiplicative inflation factor of $1.0$. To maintain ensemble spread, we inject additive system noise at each assimilation time using a precomputed spatial covariance estimated from HR training simulations; the noise magnitude is scaled by a factor of $0.005$, with a larger factor of $0.2$ in the initial cycle.

All training, evaluation, and diagnostic computations were performed on a machine equipped with an Intel Xeon Gold 6326 CPU and an NVIDIA RTX A6000 GPU.

\section{Results}
\label{sec:results}

\begin{figure}
  \centering
  \includegraphics[width=0.8\textwidth]{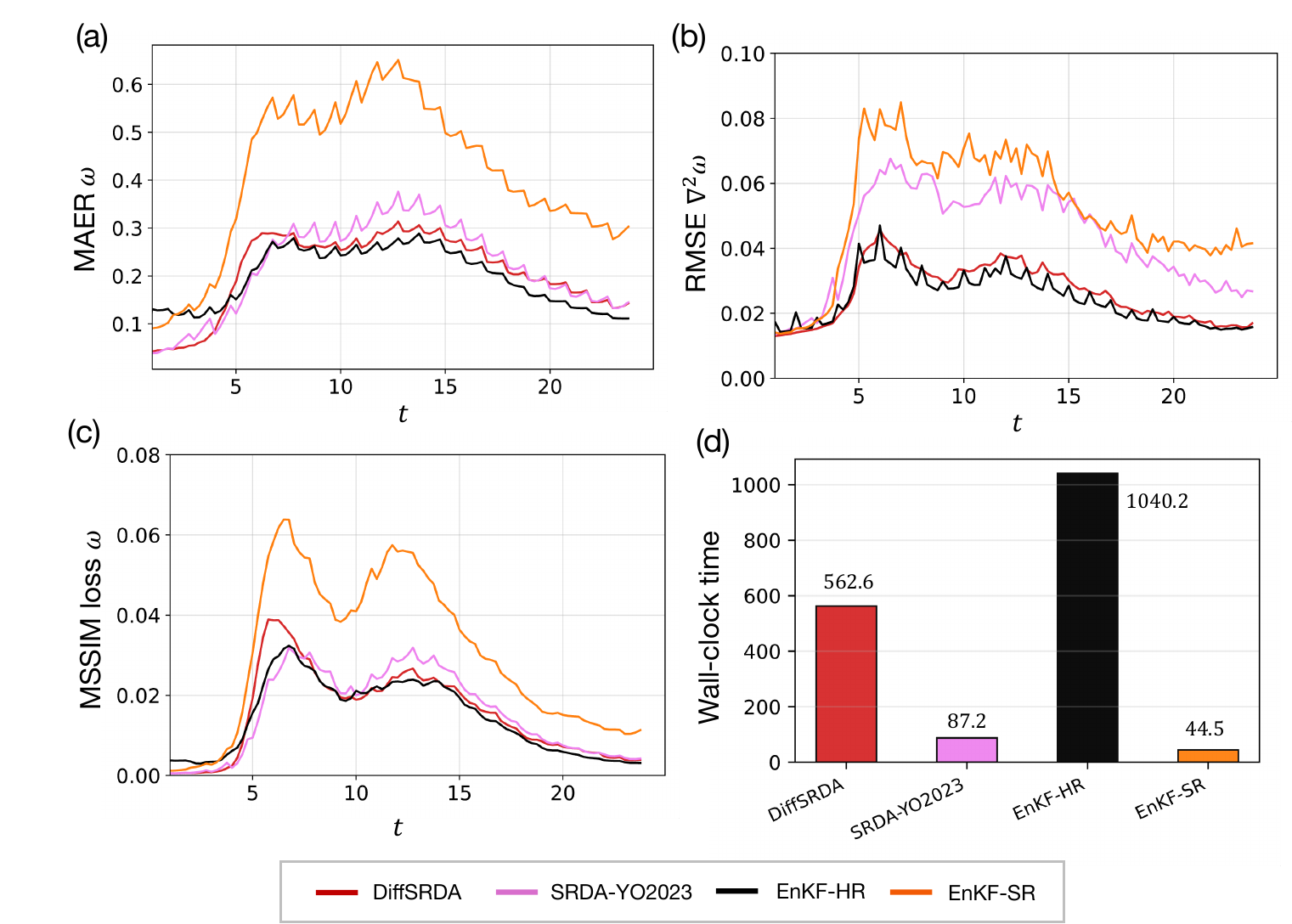}
  \caption{Time evolution of point-estimate accuracy and computational cost for the different data assimilation methods, evaluated against the UHR reference. (a) Mean absolute error ratio (MAER) of the vorticity field, (b) root-mean-square error (RMSE) of its discrete Laplacian, (c) MSSIM loss, and (d) average wall-clock time to complete one full cycling run from $t=0$ to $t=24$, measured on a machine with an Intel Xeon Gold 6326 CPU and a single NVIDIA RTX A6000 GPU. Computations are performed primarily on the GPU, with additional CPU-side overhead included in the wall-clock time. Method labels are shown in the legend.}
\label{fig:results_errormetrics_time}
\end{figure}

We first assess the point-estimate accuracy and computational cost of the proposed diffusion-based SRDA method, and then examine how the reverse-sampling budget controls the accuracy--cost trade-off. Unless stated otherwise, all error curves are averaged over 40 UHR reference simulations and are evaluated on the UHR grid after mapping HR outputs to UHR by bicubic interpolation.

Figure~\ref{fig:results_errormetrics_time} summarizes the time evolution of the accuracy metrics together with the average wall-clock time for one full cycling run from $t=0$ to $t=24$ for all methods. For DiffSRDA, the point estimate shown in panels (a)--(c) is the posterior mean, computed as the ensemble average over repeated conditional samples. In this figure, we use the full reverse chain ($T=1000$ reverse steps) in order to establish a clear upper reference for DiffSRDA in terms of point-estimate accuracy.

In terms of the vorticity error (Figure~\ref{fig:results_errormetrics_time}(a)), EnKF-HR attains the lowest mean absolute error ratio (MAER) over most of the time window, as expected since it assimilates using an HR forecast model and a large ensemble. Among the learned SRDA operators, DiffSRDA yields the smallest MAER, while the deterministic SRDA baseline of \citet{yasuda2023spatio}, which we denote hereafter by SRDA-YO2023, is less accurate. EnKF-SR, which combines LR forecasting with bicubic super-resolution inside the EnKF update, performs substantially worse in this setting. The relative gaps depend on the dynamical stage: differences are modest during the early evolution and in the late-time regime, but widen noticeably around the instability-growth and strongly nonlinear stages. The same overall ordering is reflected in the MSSIM loss in Figure~\ref{fig:results_errormetrics_time}(c), indicating that DiffSRDA better preserves coherent structures than SRDA-YO2023.

To probe fine-scale content more directly, Figure~\ref{fig:results_errormetrics_time}(b) reports the RMSE of the discrete Laplacian of the vorticity field, which strongly amplifies thin filaments and sharp gradients. Here the benefit of diffusion-based SRDA is more pronounced: DiffSRDA reduces the Laplacian error more substantially than SRDA-YO2023, suggesting improved reconstruction of small-scale structure beyond what is visible from MAER alone.

The main practical limitation of full-chain sampling appears in Figure~\ref{fig:results_errormetrics_time}(d). With $T=1000$ reverse steps, DiffSRDA is considerably more expensive over a full cycling run than the deterministic SRDA operator, although it remains substantially cheaper than EnKF-HR. This motivates the next experiment, in which we vary the number of reverse steps through timestep respacing to reduce the cost of DiffSRDA toward that of SRDA-YO2023 while aiming to retain the accuracy gains observed with full-chain sampling.

\begin{figure}
  \centering
  \includegraphics[width=0.8\textwidth]{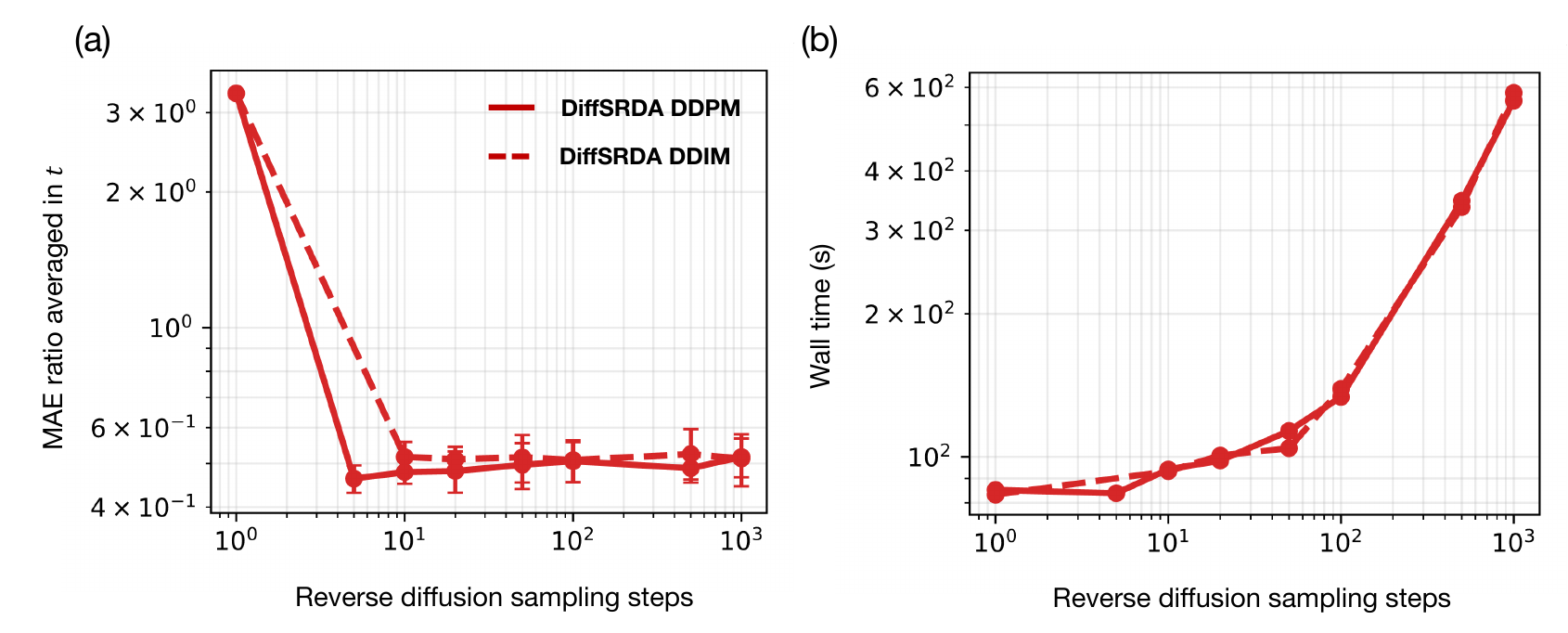}
  \caption{Effect of the number of reverse diffusion steps ($1,5,10,20,50,100,500,1000$) on accuracy and cost for DiffSRDA. (a) MAER of the vorticity field averaged over the full simulation time window, and (b) average wall-clock time for one full cycling run from $t=0$ to $t=24$. Results are shown for DDPM-style sampling ($\eta=1$) and DDIM-style sampling ($\eta=0$).}
\label{fig:results_respacing}
\end{figure}

Figure~\ref{fig:results_respacing} quantifies how accuracy and cost change with the number of reverse diffusion steps. We report the MAER averaged over the full simulation time window (Figure~\ref{fig:results_respacing}(a)) and the corresponding average wall-clock time for one full cycling run from $t=0$ to $t=24$ (Figure~\ref{fig:results_respacing}(b)). Results are shown for DiffSRDA using both DDPM-style sampling ($\eta=1$) and DDIM-style sampling ($\eta=0$).

Two trends stand out. First, extremely short reverse chains are insufficient: the one-step setting produces large errors. Second, once the reverse chain includes a small but nontrivial number of steps, the accuracy saturates rapidly. In particular, using five reverse steps yields a time-averaged MAER comparable to the 1000-step setting, and in our runs it is sometimes slightly lower. This indicates that timestep respacing can retain the benefits of diffusion-based SRDA while avoiding most of the sampling cost. The cost in Figure~\ref{fig:results_respacing}(b) drops sharply as the number of reverse steps decreases. The scaling is not perfectly linear because each full cycling run includes fixed overhead, such as conditioning preparation and model I/O, but the overall conclusion is unchanged: most of the cost of full-chain sampling can be removed without sacrificing accuracy. Based on this result, we use the five-step setting (TR5) for the qualitative comparisons below and for the uncertainty-quantification experiments that require repeated sampling. From this point onward, unless stated otherwise, DiffSRDA refers to the respaced sampler with five reverse steps (TR5).

\begin{figure}
  \centering
  \includegraphics[width=1.0\textwidth]{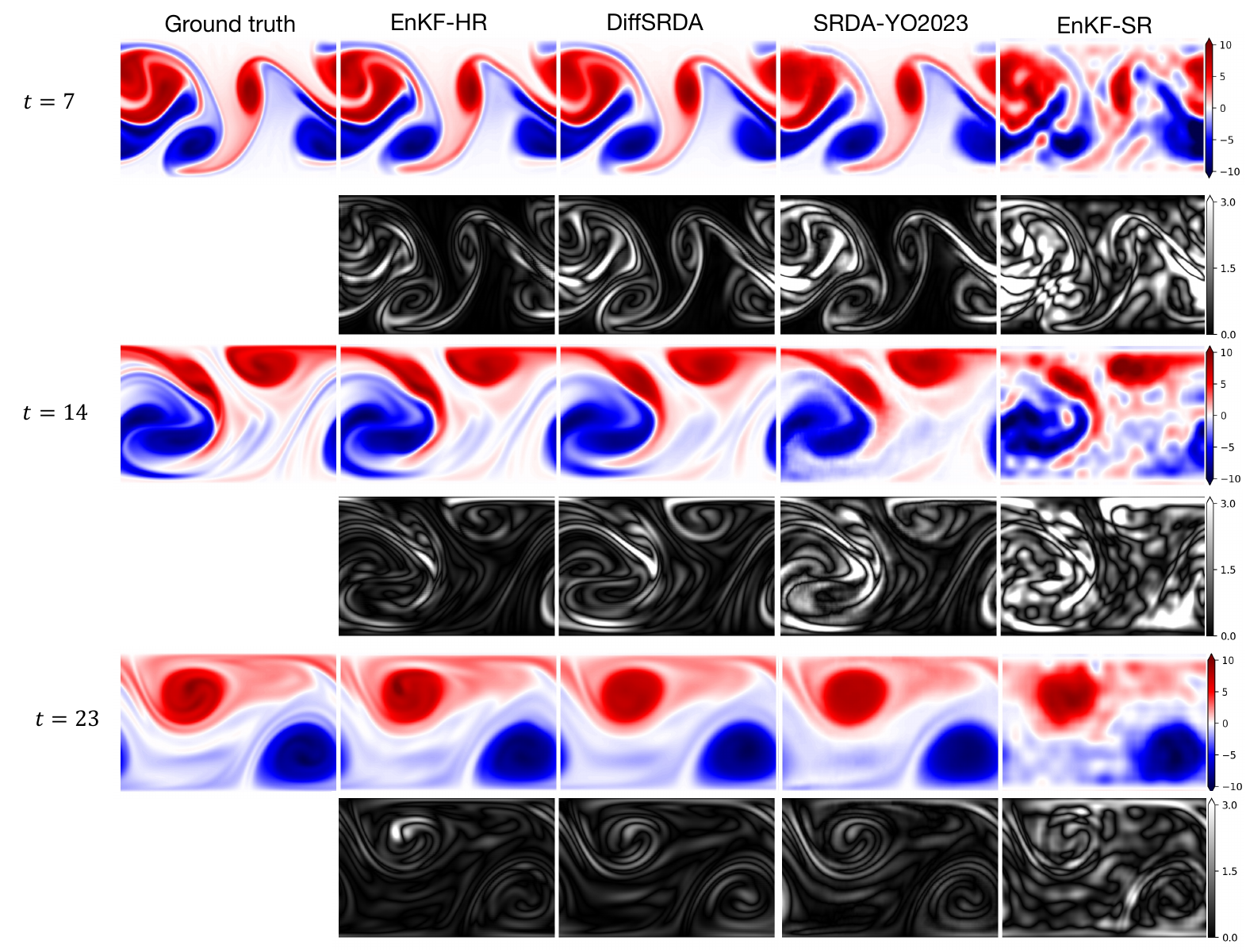}
  \caption{Vorticity snapshots at $t=7$, $t=14$, and $t=23$ comparing the UHR reference with EnKF-HR, DiffSRDA, SRDA-YO2023, and EnKF-SR. For each case, the absolute error relative to the UHR reference is shown beneath the corresponding vorticity field. The selected times represent the early meandering stage ($t=7$), the vortex-merger stage ($t=14$), and the late-time regime ($t=23$), and they coincide with prominent error levels in Figure~\ref{fig:results_errormetrics_time}. For DiffSRDA, we show the TR5 setting (five reverse diffusion steps).}
\label{fig:results_flowfield_snapshots}
\end{figure}

\begin{figure}
  \centering
  \includegraphics[width=0.8\textwidth]{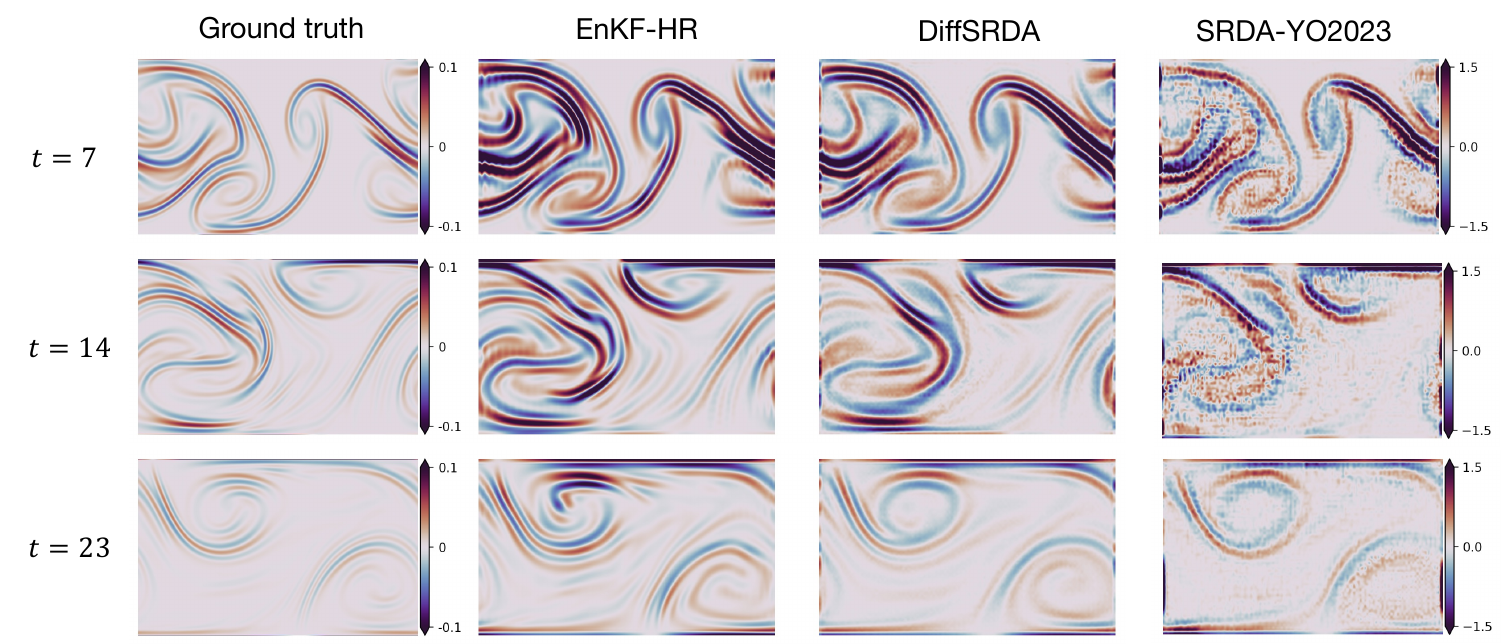}
  \caption{Snapshots of the Laplacian of the vorticity field at $t=7$, $t=14$, and $t=23$ comparing the UHR reference with EnKF-HR, DiffSRDA, and SRDA-YO2023.}
\label{fig:results_laplacian_snapshots}
\end{figure}

Figure~\ref{fig:results_flowfield_snapshots} compares vorticity snapshots and spatial error fields at three representative times, $t=7$, $t=14$, and $t=23$, corresponding to the early meandering stage, the vortex-merger stage, and the late-time regime. These times also align with prominent error levels in Figure~\ref{fig:results_errormetrics_time}. EnKF-HR provides the closest visual match to the UHR reference, consistent with its lower errors. EnKF-SR exhibits noticeably degraded structures and large errors across all three times. Among the learned SRDA frameworks, the deterministic SRDA-YO2023 baseline shows a rougher and patchier appearance with more visible small-scale noise, whereas DiffSRDA produces reconstructions that are closer to the UHR field in their large-scale vortex organization. Across all methods, the largest absolute errors concentrate near strong gradients and filamentary structures, which are the most difficult features to reconstruct from sparse observations and low-cost forecasts.

To further isolate differences in small-scale structure, Figure~\ref{fig:results_laplacian_snapshots} compares snapshots of the Laplacian of the vorticity field at the same times. This quantity is more sensitive than the vorticity field itself because it emphasizes sharp spatial variations and thin filaments, making differences easier to identify even when the vorticity fields appear broadly similar. Consistent with the Laplacian-RMSE trends in Figure~\ref{fig:results_errormetrics_time}(b), DiffSRDA most closely matches the spatial structure of the UHR Laplacian among the learned SRDA methods. SRDA-YO2023 shows more pronounced patchiness and spurious high-frequency content, reflecting its weaker reconstruction of fine-scale features.

Overall, DiffSRDA provides the strongest point-estimate accuracy among the learned SRDA operators and remains closest to the EnKF-HR baseline across the metrics considered. Most importantly, the respacing study shows that diffusion-based SRDA does not require long reverse chains in this setting: using five reverse steps retains essentially the same accuracy as $T=1000$ while reducing computational cost by a large factor, making pixel-space diffusion practical for repeated assimilation cycles. We also evaluate a latent diffusion variant as an ablation study, which is summarized in Appendix~\ref{app:SRDA_ldm}.

Beyond the empirical improvements, the diffusion formulation also offers a useful interpretation of why DiffSRDA performs well and why its accuracy can approach EnKF-HR even though it does not run an HR forecast ensemble. As discussed in Appendix~\ref{sec:bayes_posterior_connection}, the windowed conditional diffusion model can be interpreted as sampling from a Bayesian analysis distribution conditioned on two sources of information: (i) the sparse noisy observations within the window and (ii) the LR forecast trajectory used as conditioning input. Although the LR forecast is low fidelity, it still carries forward dynamical information from previous cycles, so the diffusion model implicitly inherits part of the temporal history that classical sequential DA methods obtain through repeated forecast--analysis updates. In this sense, some of the missing HR cycling information is supplied through the LR-conditioning pathway, which helps make the diffusion posterior both accurate and physically plausible rather than an unstructured black-box mapping.

Another advantage of diffusion sampling is that it avoids the familiar blurring failure mode of deterministic super-resolution regressors. Deterministic CNN-based SR models trained with regression-style losses tend to learn a single point estimate, often close to the conditional mean, so when the LR-to-HR mapping is not uniquely determined, or when the LR statistics shift, the prediction can become over-smoothed or exhibit patchy small-scale artifacts. In contrast, conditional diffusion models learn the conditional score through denoising and generate samples through iterative stochastic refinement from Gaussian noise. This process can be viewed as repeatedly moving toward higher-probability regions of the conditional data distribution in a Langevin-like manner, allowing multiple plausible HR reconstructions for the same LR input rather than collapsing to a single average field \citep{saharia2022image,song2020score}.

This probabilistic output is particularly valuable for forecasting applications because it provides not only an analysis mean but also a distribution over plausible HR analysis windows, thereby enabling uncertainty quantification through ensemble spread. In the next section, we investigate this aspect in more detail.

\section{Uncertainty quantification and calibration}
\label{sec:uq}

Because of its generative nature, a practical advantage of DiffSRDA is that it produces an ensemble of HR analyses through repeated conditional sampling, that is, by rerunning the reverse diffusion sampler with different initial Gaussian noise while keeping the same LR-forecast and observation conditions fixed. In this section, we use these ensembles to examine uncertainty characteristics and compare them with EnKF-HR, which represents uncertainty through a forecast ensemble and the Kalman update. Unless stated otherwise, DiffSRDA results use TR5 (five reverse steps).

\begin{figure}
  \centering
  \includegraphics[width=0.9\textwidth]{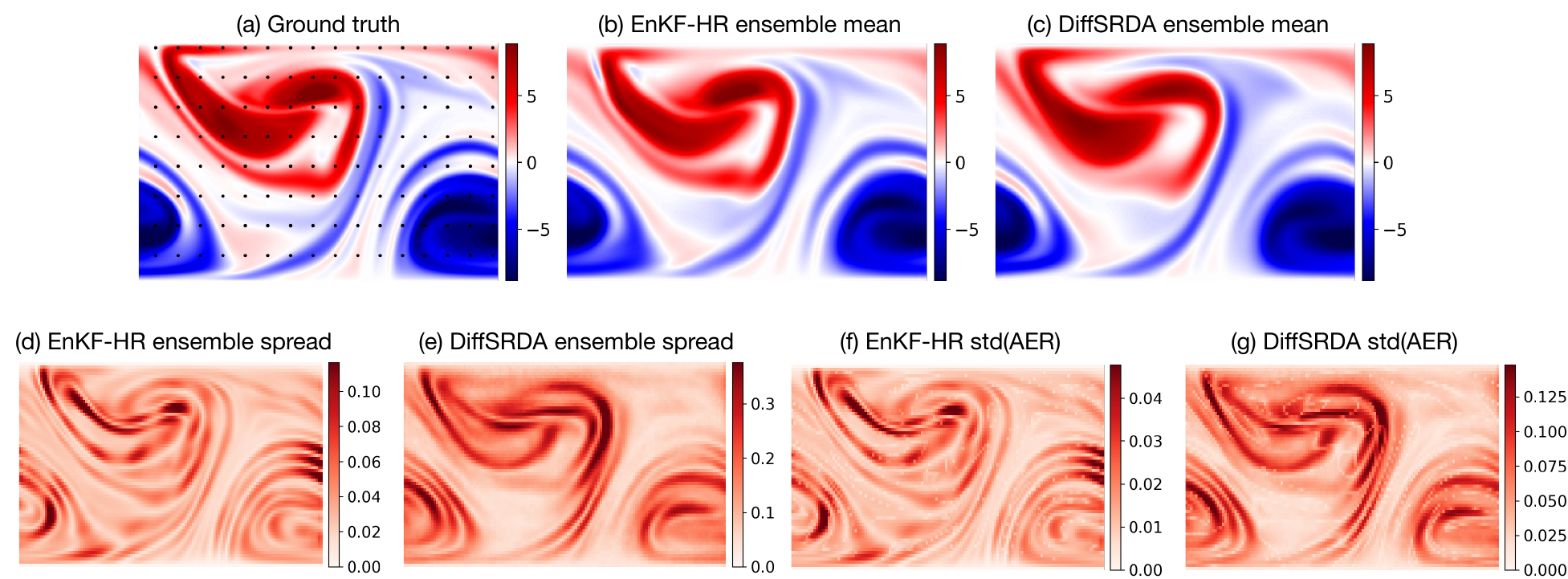}
  \caption{Ensemble mean and spread of the vorticity field at $t=16$ for (a) the HR-grid reference obtained from the UHR trajectory, (b) EnKF-HR, and (c) DiffSRDA. The bottom row shows the ensemble spread $\mathrm{Std}_b[\mathbf{x}^{(b)}]$ for (d) EnKF-HR and (e) DiffSRDA, and the spread of the absolute error $\mathrm{Std}_b[\mathrm{AER}^{(b)}]$ computed using the reference field for (f) EnKF-HR and (g) DiffSRDA. DiffSRDA results use five reverse sampling steps.}
\label{fig:uq_snapshots}
\end{figure}

\begin{figure}
  \centering
  \includegraphics[width=0.7\textwidth]{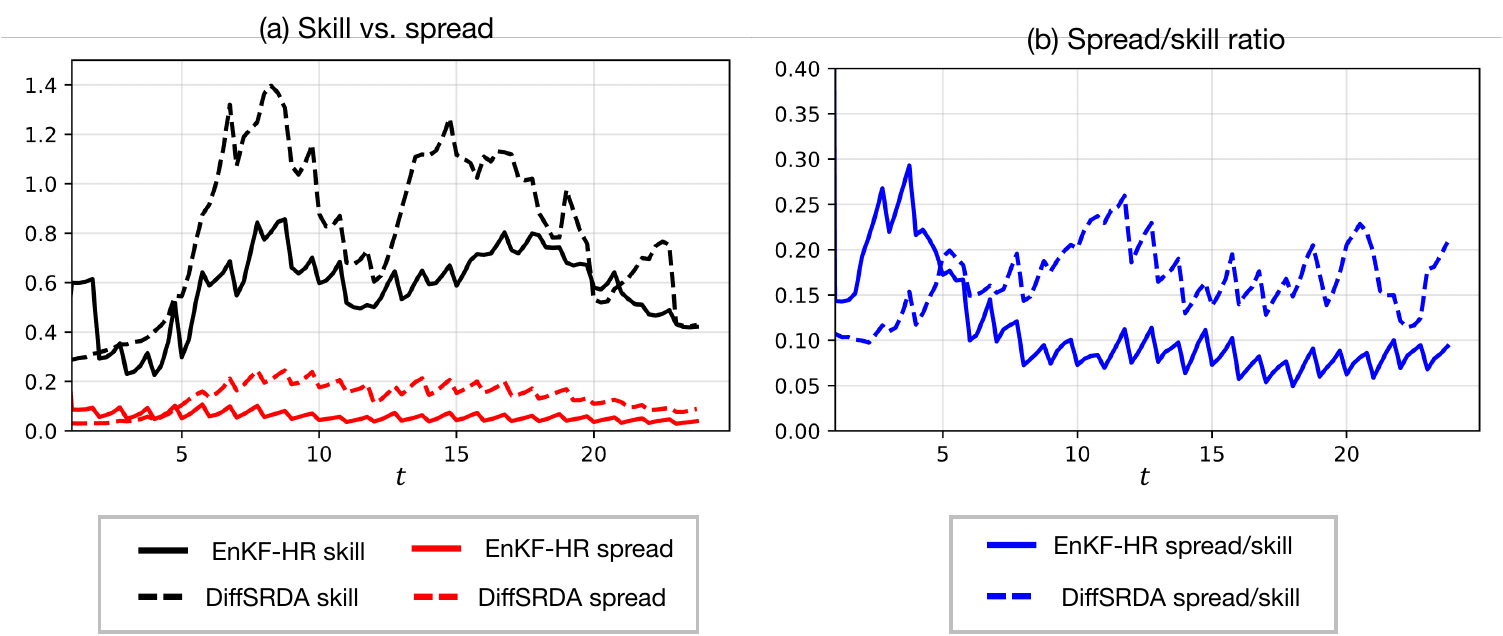}
  \caption{Time series of (a) skill, defined as the RMSE of the ensemble mean against the HR-grid reference, and spread, defined as the square root of the domain-mean ensemble variance, and (b) the spread/skill ratio for EnKF-HR and DiffSRDA.}
\label{fig:uq_skill_spread}
\end{figure}

\begin{figure}
  \centering
  \includegraphics[width=0.5\textwidth]{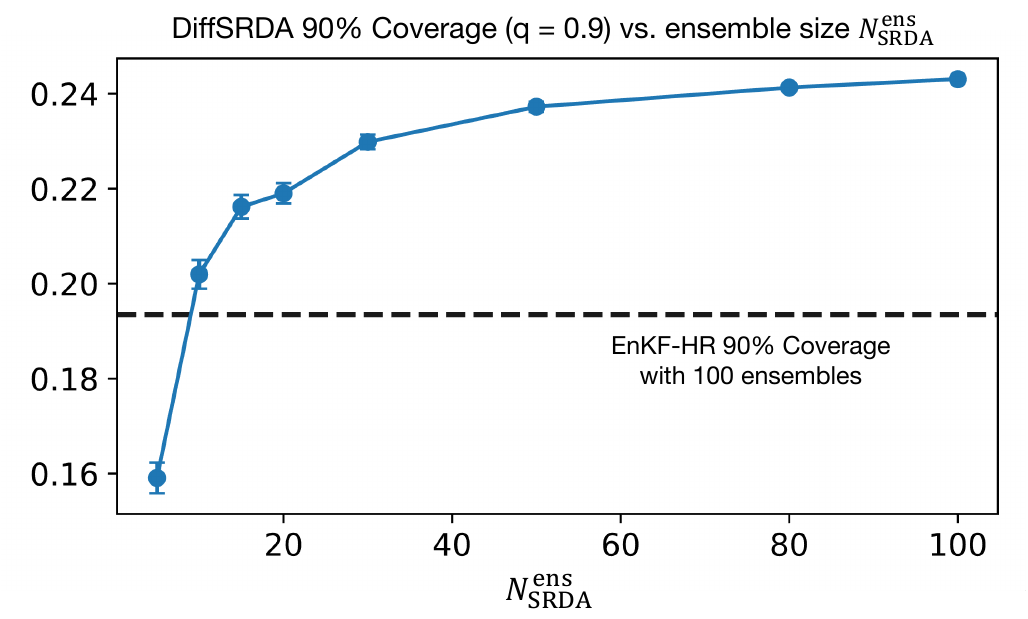}
  \caption{Sensitivity of DiffSRDA reliability to ensemble size. The $90\%$ empirical coverage ($q=0.9$) is shown as a function of the number of DiffSRDA ensemble members $N^{\mathrm{ens}}_{\mathrm{SRDA}}$. The dashed line marks the EnKF-HR $90\%$ coverage obtained with the fixed ensemble size of 100 used throughout this study. DiffSRDA coverage increases rapidly with ensemble size, exceeds the EnKF-HR level by about $N^{\mathrm{ens}}_{\mathrm{SRDA}}\approx 10$, and changes only weakly beyond $N^{\mathrm{ens}}_{\mathrm{SRDA}}\approx 30$.}
\label{fig:uq_ensemble_size}
\end{figure}

We organize the discussion around Figures~\ref{fig:uq_snapshots}--\ref{fig:uq_ensemble_size}. We first give a snapshot-level view of ensemble mean and spread (Figure~\ref{fig:uq_snapshots}), then summarize spread and skill over time (Figure~\ref{fig:uq_skill_spread}), assess reliability using coverage and rank histograms (Figure~\ref{fig:uq_coverage_rankhist}), and finally examine the sensitivity of the coverage estimate to the DiffSRDA ensemble size (Figure~\ref{fig:uq_ensemble_size}). For Figures~\ref{fig:uq_snapshots}--\ref{fig:uq_coverage_rankhist}, we use 100 ensemble members for both EnKF-HR and DiffSRDA.

The diagnostics used here are standard for evaluating whether an ensemble is not only accurate on average but also statistically consistent with its own uncertainty claims. A basic dispersion check compares the ensemble spread with the error of the ensemble mean (skill). Under ideal conditions, where the reference state and ensemble members can be viewed as draws from the same distribution and the ensemble mean is unbiased, the ensemble variance should match the mean-squared error of the ensemble mean in expectation, so an appropriately aggregated spread should agree with RMSE on average \citep{fortin2014should}. Because this equality can fail in practice for many reasons, including finite ensembles, model error, and imperfect observation handling, we also report two calibration diagnostics that act more directly on the empirical predictive distribution: (i) coverage of central prediction intervals, which should be close to the nominal level for a calibrated probabilistic forecast \citep{gneiting2005calibrated,gneiting2007strictly}, and (ii) rank histograms (Talagrand diagrams), where a roughly uniform histogram indicates reliability while systematic shapes diagnose bias and under or over-dispersion \citep{talagrand1997evaluation,hamill2001interpretation}.

As a reference for verification, we use the HR field obtained by downsampling the UHR trajectory to the HR grid. Let $\mathbf{x}^{(b)}_t$ denote the HR vorticity field of ensemble member $b$ at physical time $t$, and let $\mathbf{x}^{\mathrm{ref}}_t$ denote the corresponding HR-grid reference field. The ensemble mean and pointwise ensemble spread are defined as
\[
\mu_{t,i}=\mathbb{E}_b\!\left[[\mathbf{x}^{(b)}_t]_i\right],
\qquad
\sigma_{t,i}=\mathrm{Std}_b\!\left([\mathbf{x}^{(b)}_t]_i\right),
\]
where $i$ indexes HR grid points.

Figure~\ref{fig:uq_snapshots} provides a compact snapshot-level view of ensemble mean and spread at $t=16$. The top row compares the reference field with the EnKF-HR and DiffSRDA ensemble means. Consistent with the point-estimate results, the EnKF-HR mean is typically closer to the reference, while DiffSRDA shows slightly larger deviations in regions with sharp gradients and filamentary structure. Panels (d,e) show the analysis spread $\sigma_{t,i}$. A key observation is that the DiffSRDA spread pattern is qualitatively similar to that of EnKF-HR: both methods place their largest uncertainty in the same dynamically active regions where gradients are strongest. DiffSRDA produces a broader spread overall than EnKF-HR, but the spatial pattern of uncertainty closely matches that of the EnKF-HR baseline.

Panels (f,g) show $\mathrm{Std}_b[\mathrm{AER}^{(b)}_{t,i}]$, the standard deviation across ensemble members of a normalized absolute error computed against the reference, where
\[
\mathrm{AER}^{(b)}_{t,i}
=
\frac{
\left|[\mathbf{x}^{(b)}_t]_i-[\mathbf{x}^{\mathrm{ref}}_t]_i\right|
}{
\mathbb{E}_{i}\!\left[\left|[\mathbf{x}^{\mathrm{ref}}_t]_i\right|\right]
}.
\]
Because this quantity is computed using the reference field, it is not an operational uncertainty estimate. We include it only as a diagnostic to highlight where ensemble-member errors vary most strongly across the domain.

Figure~\ref{fig:uq_skill_spread} summarizes ensemble behavior over time using two scalar curves. We define the skill as the RMSE of the ensemble mean against the reference,
\begin{equation}
\mathrm{Skill}(t)=
\sqrt{
\mathbb{E}_{i}\!\left[
\left(\mu_{t,i}-[\mathbf{x}^{\mathrm{ref}}_t]_i\right)^2
\right]
},
\label{eq:skill_def}
\end{equation}
and the spread as the square root of the domain-mean ensemble variance \citep{fortin2014should},
\begin{equation}
\mathrm{Spread}(t)=
\sqrt{
\mathbb{E}_{i}\!\left[\sigma_{t,i}^2\right]
}.
\label{eq:spread_def}
\end{equation}
Skill measures the accuracy of the ensemble mean, whereas spread measures how widely the ensemble members vary around that mean. In Figure~\ref{fig:uq_skill_spread}(a), EnKF-HR attains lower skill, meaning better mean accuracy, over most times, while DiffSRDA exhibits a larger spread, consistent with the results so far. Figure~\ref{fig:uq_skill_spread}(b) reports the ratio $\mathrm{Spread}(t)/\mathrm{Skill}(t)$ as a simple indicator of how the ensemble width compares with the typical mean error. In our setting, this ratio stays well below 1 for both methods over most times, which is consistent with under-dispersion under a strict interpretation. DiffSRDA tends to have a larger spread/skill ratio than EnKF-HR, indicating that its ensemble is less concentrated even though its mean error is slightly larger.

\begin{figure}
  \centering
  \includegraphics[width=0.7\textwidth]{figures/uq_coverage_rankhist.pdf}
  \caption{Reliability diagnostics for ensemble uncertainty. (a) Empirical coverage versus time for nominal central intervals $(50\%, 80\%, 90\%)$ for EnKF-HR (solid) and DiffSRDA (dashed). (b,c) Rank histograms of the HR-grid reference within the ensemble at the pixel level for (b) EnKF-HR and (c) DiffSRDA.}
\label{fig:uq_coverage_rankhist}
\end{figure}

Because the spread/skill comparison is only a partial dispersion check for a spatially correlated field, we also assess pixelwise calibration using empirical coverage and rank histograms in Figure~\ref{fig:uq_coverage_rankhist}. Figure~\ref{fig:uq_coverage_rankhist}(a) shows the empirical coverage of central prediction intervals for nominal levels $q\in\{0.5,0.8,0.9\}$. Coverage is a direct calibration check for predictive intervals: for a calibrated probabilistic forecast, the empirical fraction of events falling inside a nominal $q$ interval should be close to $q$ \citep{gneiting2005calibrated,gneiting2007strictly}. For each pixel, we compute ensemble quantiles and then report the fraction of pixels whose reference value lies within the nominal interval. If the ensemble were well calibrated at the pixel level, the coverage would be close to $q$. In our results, both methods show coverage well below nominal, especially for the 90\% interval, indicating substantial under-coverage. DiffSRDA achieves higher coverage than EnKF-HR under the same observation setting, consistent with its larger spread.

Figure~\ref{fig:uq_coverage_rankhist}(b,c) shows rank histograms (Talagrand diagrams), constructed by ranking the reference value among the sorted ensemble members at each pixel. A roughly flat histogram is consistent with reliability, while a U-shaped histogram indicates under-dispersion, meaning that the reference often falls outside the ensemble range, and a sloped histogram indicates bias. The results show that both methods exhibit strongly U-shaped histograms, again indicating under-dispersion, with DiffSRDA being slightly less extreme, consistent with its higher empirical coverage.

These diagnostics verify ensembles against a higher-fidelity reference while both methods produce HR analyses constrained by HR-resolution observations, so the absolute calibration levels should be interpreted cautiously. The main point here is comparative: under the same observation setting and verification procedure, DiffSRDA yields a broader spread and correspondingly improved reliability diagnostics relative to EnKF-HR.

Finally, Figure~\ref{fig:uq_ensemble_size} examines how many DiffSRDA samples are needed for stable reliability estimates. We report the mean 90\% coverage as a function of the DiffSRDA ensemble size $N^{\mathrm{ens}}_{\mathrm{SRDA}}$, computed by subsampling from a larger pool of samples. Coverage increases rapidly for small ensembles and changes only weakly beyond roughly $N^{\mathrm{ens}}_{\mathrm{SRDA}}\approx 30$. The dashed line marks the EnKF-HR 90\% coverage obtained with its fixed ensemble size of 100 used throughout this study. In this setting, DiffSRDA reaches comparable or higher coverage with on the order of 10 samples and shows diminishing returns beyond about 30. This experiment is best viewed as a Monte Carlo convergence check: increasing $N^{\mathrm{ens}}_{\mathrm{SRDA}}$ improves estimates of the ensemble mean, quantiles, and coverage, but it does not by itself remove the under-coverage seen in Figure~\ref{fig:uq_coverage_rankhist}. The practical implication is that DiffSRDA uncertainty diagnostics can be estimated with a modest number of conditional samples, which keeps probabilistic SRDA computationally feasible even when large forecast ensembles are expensive.

Taken together, Figures~\ref{fig:uq_snapshots}--\ref{fig:uq_ensemble_size} indicate that DiffSRDA produces a clear and physically meaningful uncertainty signal, with spread patterns that closely mirror those of EnKF-HR, and that its reliability diagnostics are comparable to, and in some cases slightly better than, those of EnKF-HR under the same observation setting.

\section{Deployment-time training-free observation-consistency guidance}
\label{sec:obsguid}

In this section, we demonstrate how the score-based structure of DiffSRDA enables \emph{training-free observation-consistency guidance} at deployment time. Up to this point, DiffSRDA has been trained and evaluated using sparse mask-type observations on the HR grid with observation grid interval 8 (\texttt{ogi8}). We now consider deployment-time shifts in the observation configuration while keeping the trained diffusion model fixed. Two cases are of particular interest. The first is a structured sensor-improvement scenario in which the observation grid becomes denser, here moving from interval 8 to interval 4 (\texttt{ogi4}). The second is a less structured deployment shift in which the available sensors occupy fixed but spatially random locations rather than the regular lattice used during training. In both situations, directly feeding the new observations through the original conditioning interface may induce a distribution shift and degrade reconstruction quality, whereas retraining the conditional diffusion model for each new sensor layout is effective but expensive. Our goal is to show that guidance provides a simple deployment-time alternative by incorporating the new observations through an explicit observation-consistency correction during reverse diffusion sampling, without modifying the trained network.

This approach follows the standard Bayesian perspective used in many diffusion-based inverse problems, including super-resolution, sparse reconstruction, and inpainting, where a pretrained generative model acts as a prior and measurement information is incorporated during inference. In score-based diffusion models, the network provides a multi-noise-level prior score, and observation information can be injected by modifying the reverse-time sampling dynamics \citep{song2020score,ho2020denoising}. We refer to \citet{daras2024survey,chung2025diffusion} for broader reviews and practical guidance methods, including inpainting-style corrections such as RePaint \citep{lugmayr2022repaint}, gradient-based posterior sampling methods such as diffusion posterior sampling (DPS) \citep{chung2022diffusion}, and more recent refinements aimed at improving robustness \citep{zhang2025improving}. More specifically related to data assimilation, \citet{rozet2023score} discusses guidance strategies tailored to score-based DA.

Let $\mathbf{x}$ denote a single HR state and let the SRDA output be the five-frame HR analysis window $\mathbf{X}=[\mathbf{x}^{\mathrm{HR}}_t,\ldots,\mathbf{x}^{\mathrm{HR}}_{t+4}]$. Let $\mathbf{c}$ denote the conditioning inputs used by the trained DiffSRDA model, namely the LR forecast information together with the processed observation channels provided to the network. Without guidance, the model defines an amortized conditional distribution $p_\theta(\mathbf{X}_0\mid \mathbf{c})$ through the learned reverse diffusion chain.
At inference time, suppose we additionally obtain observations $\mathbf{y}$ on the HR grid. In an idealized setting, these observations may provide information that is not already represented in $\mathbf{c}$ or that falls outside the training observation distribution. We then consider the guided target
\begin{equation}
\tilde p(\mathbf{X}_0\mid \mathbf{c},\mathbf{y})
\propto
p_\theta(\mathbf{X}_0\mid \mathbf{c})\,p(\mathbf{y}\mid \mathbf{X}_0).
\label{eq:sec7_guided_posterior}
\end{equation}
Equation~\eqref{eq:sec7_guided_posterior} is a useful idealized target for motivating the guidance construction. In the experiments reported below, however, guidance is best interpreted as an observation-consistency correction applied to an already observation-conditioned DiffSRDA model, rather than as exact Bayesian conditioning on entirely new information beyond the conditioning channels.

At diffusion step $\tau$, the corresponding guided score admits the decomposition
\begin{equation}
\nabla_{\mathbf{X}_\tau}\log \tilde p_\tau(\mathbf{X}_\tau\mid \mathbf{c},\mathbf{y})
=
\nabla_{\mathbf{X}_\tau}\log p_{\theta,\tau}(\mathbf{X}_\tau\mid \mathbf{c})
+
\nabla_{\mathbf{X}_\tau}\log p_{\theta,\tau}(\mathbf{y}\mid \mathbf{X}_\tau,\mathbf{c}),
\label{eq:sec7_score_decomp_exact}
\end{equation}
where $p_{\theta,\tau}(\mathbf{y}\mid \mathbf{X}_\tau,\mathbf{c})$ denotes the model-induced diffusion-time conditional likelihood of $\mathbf{y}$ given $\mathbf{X}_\tau$ under the conditioning information $\mathbf{c}$. In practice, evaluating $\nabla_{\mathbf{X}_\tau}\log p_{\theta,\tau}(\mathbf{y}\mid \mathbf{X}_\tau,\mathbf{c})$ exactly is intractable. Following the DPS-style approximation \citep{chung2022diffusion}, we instead evaluate the likelihood at a denoised estimate of the clean state,
\begin{equation}
\nabla_{\mathbf{X}_\tau}\log p_{\theta,\tau}(\mathbf{y}\mid \mathbf{X}_\tau,\mathbf{c})
\approx
\nabla_{\mathbf{X}_\tau}\log p\!\left(\mathbf{y}\mid \hat{\mathbf{X}}_0(\mathbf{X}_\tau,\tau,\mathbf{c})\right),
\label{eq:sec7_score_decomp}
\end{equation}
where $\hat{\mathbf{X}}_0(\mathbf{X}_\tau,\tau,\mathbf{c})$ is the clean-state estimate implied by the denoiser output at step $\tau$. The full derivation, including the approximations used in our implementation, is given in Appendix~\ref{app:likelihood_guidance_derivation}. Here we summarize only the masked-residual form used in the experiments. In our implementation, the conditioning channels contain observations from the short SRDA context window at the previous and current assimilation times, whereas the guidance correction is applied only to the current analysis frame.

As in the rest of the paper, we assume a Gaussian observation model at the clean HR state,
\begin{equation}
\mathbf{y} = H\mathbf{X}_0 + \boldsymbol{\eta},
\qquad
\boldsymbol{\eta}\sim\mathcal{N}(0,R).
\label{eq:sec7_obs_model}
\end{equation}
For the mask-type on-grid operator used here, it is convenient to embed observations into the full tensor space as $\tilde{\mathbf{y}} := H^\top \mathbf{y}$ and define the binary mask matrix $M := H^\top H$. With the common simplification $R=\sigma_{\mathrm{obs}}^2 I$ on observed entries, the likelihood contribution reduces to a masked residual evaluated at the denoised estimate,
\begin{equation}
M\odot\left(\tilde{\mathbf{y}}-\hat{\mathbf{X}}_0\right).
\label{eq:sec7_mask_residual}
\end{equation}
Accordingly, we implement guidance as a correction applied to $\hat{\mathbf{X}}_0$ inside the sampler,
\begin{equation}
\hat{\mathbf{X}}_0
\leftarrow
\hat{\mathbf{X}}_0
+
\kappa_{\tau}\,\bigl(M\odot(\tilde{\mathbf{y}}-\hat{\mathbf{X}}_0)\bigr),
\label{eq:sec7_clean_update}
\end{equation}
where $\kappa_\tau$ is a guidance strength that absorbs schedule-dependent factors and likelihood scaling. In our implementation, this form is obtained via a stop-gradient, first-order Jacobian simplification (Appendix~\ref{app:likelihood_guidance_derivation}), which avoids backpropagating through the denoiser and is particularly natural for the linear mask operator used here.

Hard pointwise constraints from sparse masks can introduce grid-imprint artifacts near observed locations. To reduce this effect, we optionally smooth the masked residual with a Gaussian kernel \citep{zhang2021deep,chung2023parallel,bradbury2025your,daras2024survey},
\begin{equation}
\hat{\mathbf{X}}_0
\leftarrow
\hat{\mathbf{X}}_0
+
\kappa_{\tau}\,\Bigl(G_{\sigma_{\mathrm{blur}}} * \bigl(M\odot(\tilde{\mathbf{y}}-\hat{\mathbf{X}}_0)\bigr)\Bigr),
\label{eq:sec7_clean_update_blur}
\end{equation}

After modifying $\hat{\mathbf{X}}_0$, we can optionally recompute the corresponding noise estimate to keep the sampler algebra consistent (Appendix~\ref{app:likelihood_guidance_derivation}). Unless stated otherwise, the guided results in this section use DDPM sampling ($\eta=1$) with the blurred residual update, $T_R=20$, constant $\kappa_\tau=0.5$, $\sigma_{\mathrm{blur}}=1.5$, and no recomputation of $\boldsymbol{\varepsilon}_{\mathrm{rec}}$ after the guidance edit.

We organize the results in three stages. First, we consider a \emph{structured} deployment shift in which the model is trained with the regular \texttt{ogi8} observation layout but is deployed for inference with the denser regular \texttt{ogi4} layout. This isolates the effect of changing sensor density while preserving the same lattice structure. Second, we consider a more realistic \emph{random-layout} deployment shift in which the observation locations are fixed but spatially random rather than arranged on the regular lattices seen during training. In these random-layout experiments, we use $N_s=128$ and $N_s=512$ sensors, which match the number of observed grid points in the regular \texttt{ogi8} and \texttt{ogi4} layouts, respectively. This allows us to separate the effect of sensor count from the effect of spatial arrangement. Finally, after discussing reconstruction accuracy under these deployment shifts, we examine how guidance affects the ensemble spread and its qualitative uncertainty characteristics.

\begin{figure}
  \centering
  \includegraphics[width=1.0\textwidth]{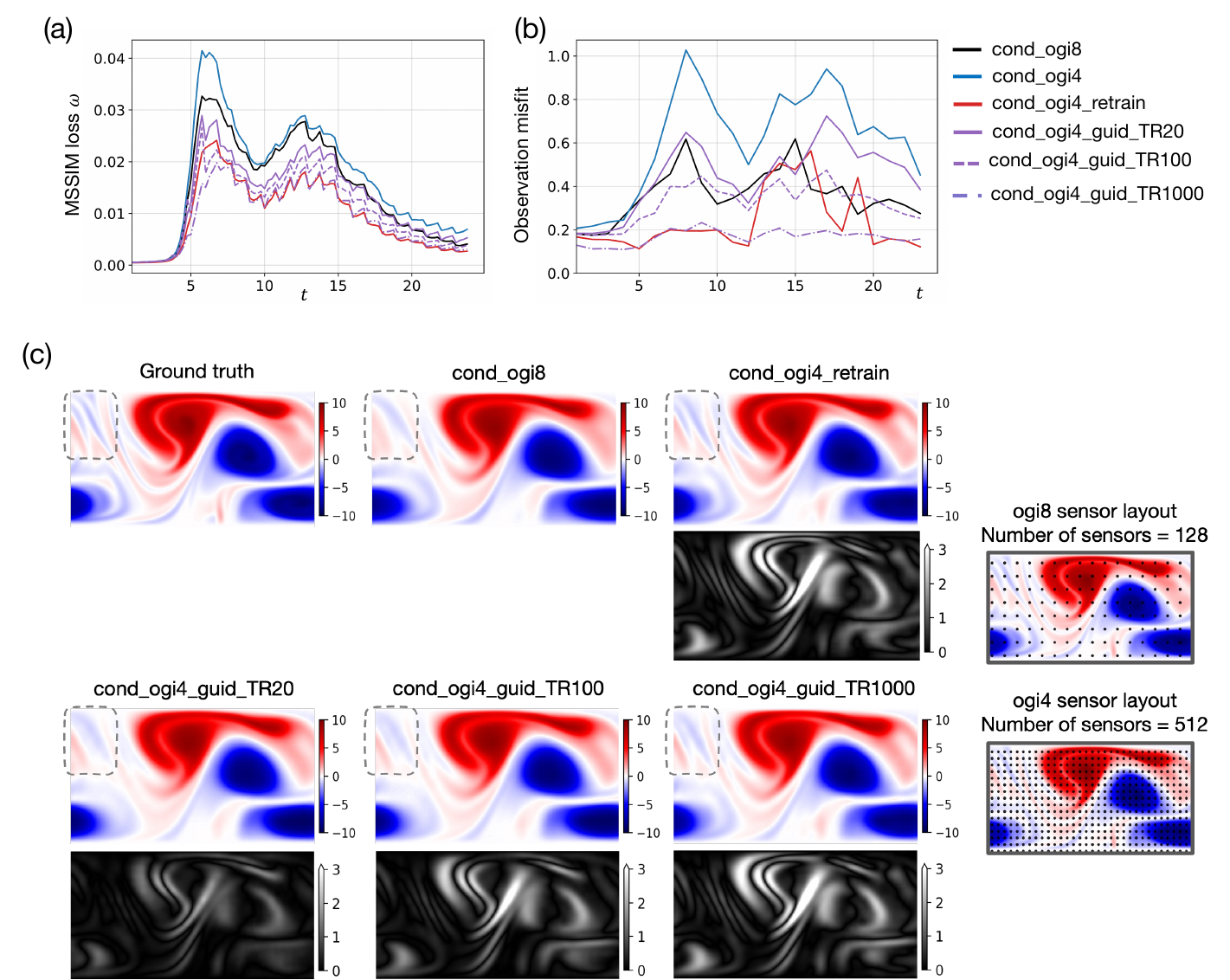}
  \caption{Effect of training-free observation-consistency guidance under a structured denser-sensor deployment shift from \texttt{ogi8} to \texttt{ogi4}. (a) MSSIM loss versus time for the different conditioning and guidance settings listed in the legend. (b) Corresponding observation misfit computed only at truly observed pixels. (c) Vorticity snapshots at $t=13$ together with grayscale maps showing the pixelwise absolute change relative to the baseline \texttt{cond\_ogi8}. The sensor-layout panels on the right show that \texttt{ogi8} and \texttt{ogi4} correspond to 128 and 512 observed grid points, respectively. See the main text for the expanded abbreviations.}
\label{fig:obsguid_results}
\end{figure}

For brevity, we use the following abbreviations. Here, \texttt{ogi$k$} denotes the regular observation grid interval $k$, so smaller $k$ means denser observations. \texttt{cond\_ogi$k$} indicates the observation pattern supplied through the conditioning channels $\mathbf{c}$, while \texttt{guid} indicates that guidance is additionally applied during sampling:
\begin{itemize}
\item \texttt{cond\_ogi8}: baseline DiffSRDA, trained and evaluated with interval-8 observations, without guidance.
\item \texttt{cond\_ogi4\_retrain}: retrained model using interval-4 observations in the conditioning channels, without guidance.
\item \texttt{cond\_ogi4}: interval-4 observations are supplied through the conditioning channels at inference time using the model trained with interval-8 conditioning, without guidance.
\item \texttt{cond\_ogi4\_guid\_TR20}, \texttt{cond\_ogi4\_guid\_TR100}, \texttt{cond\_ogi4\_guid\_TR1000}: same as \texttt{cond\_ogi4}, but with guidance applied during sampling using $TR=20,100,1000$ reverse steps, respectively.
\item \texttt{rand\_Ns128\_WithoutGuidance}: deployment with a fixed random sparse sensor layout using $N_s=128$ sensors, without guidance.
\item \texttt{rand\_Ns128\_WithGuidance}: same random-layout case with guidance applied during sampling.
\item \texttt{rand\_Ns512\_WithoutGuidance}: deployment with a fixed random sparse sensor layout using $N_s=512$ sensors, without guidance.
\item \texttt{rand\_Ns512\_WithGuidance}: same random-layout case with guidance applied during sampling.
\end{itemize}
In the random-layout experiments, all guided cases use $TR=1000$ reverse steps.

We first examine the structured deployment shift from \texttt{ogi8} to \texttt{ogi4}. Figure~\ref{fig:obsguid_results}(a,b) shows two main effects. First, naively switching to denser observations in the conditioning channels without retraining (\texttt{cond\_ogi4}) degrades performance slightly relative to the in-distribution baseline (\texttt{cond\_ogi8}). This indicates that directly injecting a new observation layout through the original conditioning interface creates a distribution shift: the model was trained to interpret interval-8 sparsity patterns and does not automatically generalize perfectly when the regular sensor lattice becomes denser. The guided runs should therefore be interpreted as applying an additional observation-consistency correction on top of an already observation-conditioned model, rather than as introducing entirely separate new information beyond the conditioning channels.

Second, incorporating the denser observations through guidance substantially improves both reconstruction quality and agreement with the observations, without retraining the network. In particular, \texttt{cond\_ogi4\_guid\_TR20} reduces MSSIM loss and lowers the observation misfit relative to \texttt{cond\_ogi4}, showing that the guided samples better match the provided measurements at observed pixels while also improving the reconstruction against the reference. Increasing the reverse-step budget strengthens this effect: moving from $TR=20$ to $TR=100$ further reduces both MSSIM loss and observation misfit, and with $TR=1000$ the guided curves approach the retrained reference \texttt{cond\_ogi4\_retrain}. This behavior is consistent with guidance acting as a small likelihood-based correction that is applied repeatedly along the reverse chain, so that larger reverse-step budgets provide more opportunities for the observation information to accumulate smoothly. Since all results here use DDPM sampling, stochastic noise is injected at every reverse step, which can help prevent overly rigid updates when the correction is applied repeatedly. These gains, however, come at increased sampling cost because larger $TR$ directly increases the number of reverse evaluations per SRDA cycle.

Figure~\ref{fig:obsguid_results}(c) complements these scalar trends with representative flow fields at $t=13$. Relative to the baseline \texttt{cond\_ogi8}, the retrained interval-4 model \texttt{cond\_ogi4\_retrain} exhibits structured changes concentrated along filamentary, high-gradient regions, which are precisely the parts of the flow that tend to dominate perceptual metrics. The guided cases \texttt{cond\_ogi4\_guid\_TR20}, \texttt{cond\_ogi4\_guid\_TR100}, and \texttt{cond\_ogi4\_guid\_TR1000} show the same overall pattern of corrections, with their magnitude and spatial coherence increasing with the reverse-step budget. The dashed inset indicates one representative region where guidance improves the reconstruction of fine filament-like structure and moves it closer to the retrained case. In addition to sharpening or adjusting filaments, some of the broader high-magnitude grayscale regions are also consistent with small spatial displacements of coherent vortical features, suggesting that guidance may correct not only local texture but also the position of larger structures. Overall, Figure~\ref{fig:obsguid_results} shows that guidance does not merely improve scalar error metrics; it makes targeted, physically interpretable adjustments in dynamically active regions of the flow, and these adjustments become progressively closer to the retrained reference as $TR$ increases.

\begin{figure}
  \centering
  \includegraphics[width=1.0\textwidth]{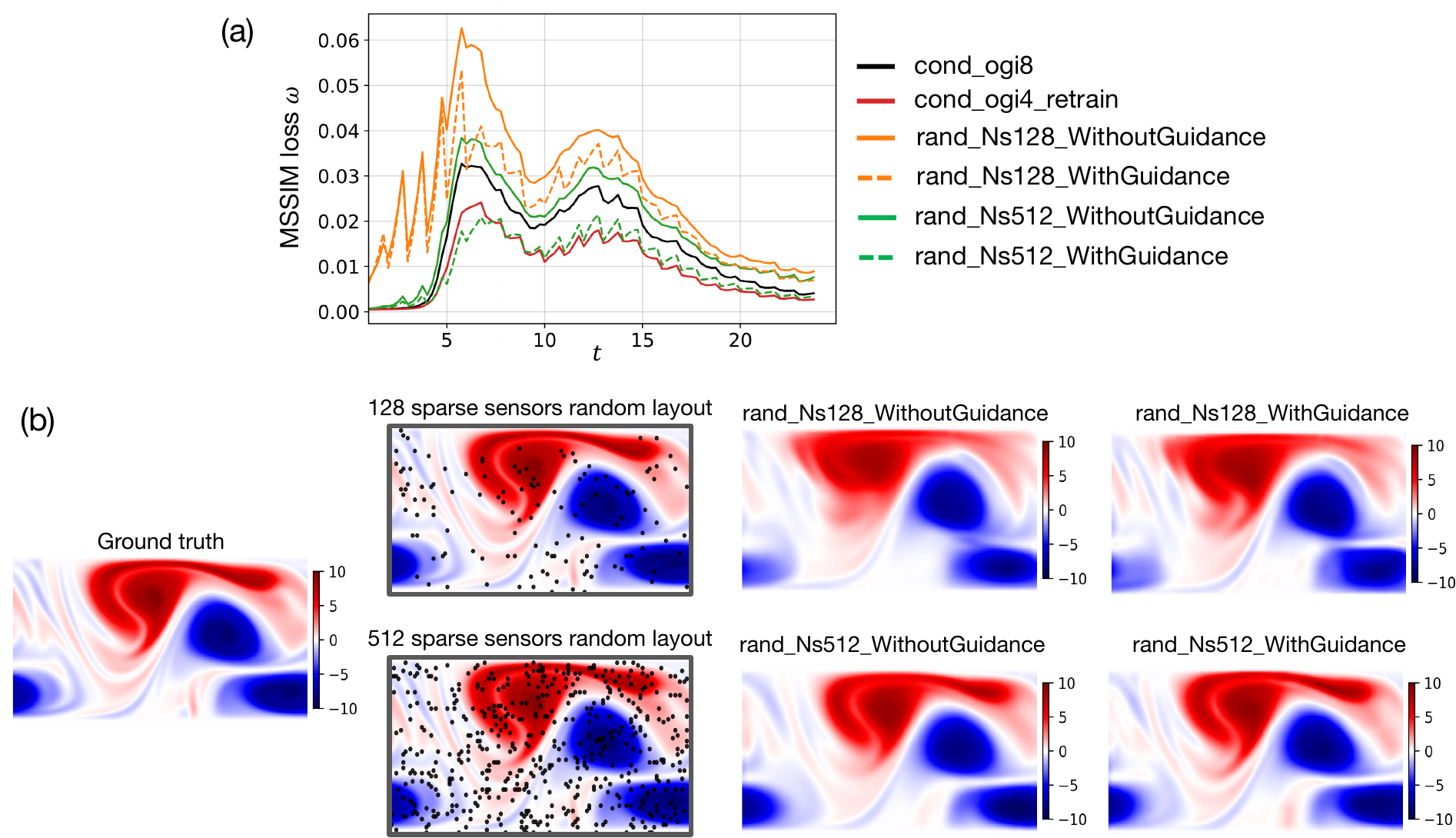}
  \caption{Effect of guidance under deployment-time shifts to fixed random sparse sensor layouts. (a) MSSIM loss versus time comparing the regular training-layout baseline \texttt{cond\_ogi8}, the retrained regular-lattice reference \texttt{cond\_ogi4\_retrain}, and the random-layout cases \texttt{rand\_Ns128\_WithoutGuidance}, \texttt{rand\_Ns128\_WithGuidance}, \texttt{rand\_Ns512\_WithoutGuidance}, and \texttt{rand\_Ns512\_WithGuidance}. Here $N_s=128$ and $N_s=512$ match the number of observed grid points in the regular \texttt{ogi8} and \texttt{ogi4} layouts, respectively. (b) Representative snapshots at $t=13$ showing the ground truth, the corresponding random sensor layouts, and the no-guidance and guided reconstructions for $N_s=128$ and $N_s=512$. All guided random-layout cases use $TR=1000$ reverse steps.}
\label{fig:obsguid_random}
\end{figure}

Figure~\ref{fig:obsguid_random} broadens the deployment study beyond regular lattices. In these experiments, the model is still the one trained under the regular \texttt{ogi8} observation pattern, but the deployment observations are now provided at fixed random sparse spatial locations rather than on a structured grid. We consider two sensor counts: $N_s=128$, which matches the number of observed points in \texttt{ogi8}, and $N_s=512$, which matches \texttt{ogi4}. This allows us to assess whether the pretrained diffusion model remains useful when the observation layout changes not only in density but also in spatial arrangement.

Several points are worth noting. First, the random-layout cases without guidance remain qualitatively reasonable, which suggests that DiffSRDA has some inherent tolerance to moderate sensor-layout shifts even when they are not seen during training. At the same time, \texttt{rand\_Ns128\_WithoutGuidance} is slightly inferior to the in-distribution regular-layout baseline \texttt{cond\_ogi8}, indicating a residual out-of-distribution penalty due to the change from a regular lattice to a random spatial arrangement. Guidance improves the reconstruction in both random-layout cases, as seen both in the MSSIM curves and in the representative snapshots in Figure~\ref{fig:obsguid_random}(b). The improvement is more pronounced for the denser $N_s=512$ case, which is natural because the larger number of measurements provides more information for the posterior correction. In fact, the performance of \texttt{rand\_Ns512\_WithGuidance} comes reasonably close to the retrained regular-lattice reference \texttt{cond\_ogi4\_retrain}, suggesting that the increase from 128 to 512 sensors partly compensates for the out-of-distribution effect of the random spatial layout. By contrast, both \texttt{rand\_Ns128\_WithoutGuidance} and \texttt{rand\_Ns128\_WithGuidance} remain somewhat worse than the structured \texttt{cond\_ogi8} case, consistent with the fact that here the sensor count is unchanged and only the off-training-distribution spatial arrangement is altered. Overall, this experiment shows that guidance is not limited to regular-lattice densification, but can also help adapt a pretrained model to more general deployment-time sensor-layout shifts.

We summarize the accuracy--cost trade-off in the structured denser-observation setting using a ``gap closed'' fraction based on a time-averaged error metric $E$, here chosen as the time-averaged MAER:
\begin{equation}
\mathrm{GapClosed}
=
\frac{E_{\mathrm{base}}-E_{\mathrm{plugin}}}{E_{\mathrm{base}}-E_{\mathrm{retrain}}}\times 100\%,
\label{eq:sec7_gapclosed}
\end{equation}
where $E_{\mathrm{base}}$ corresponds to \texttt{cond\_ogi8}, $E_{\mathrm{retrain}}$ corresponds to \texttt{cond\_ogi4\_retrain}, and $E_{\mathrm{plugin}}$ is one of the guided cases. Table~\ref{tab:sec7_gapclosed} quantifies the same trend: increasing $TR$ systematically closes more of the retraining gap, reaching near-complete closure for $TR=1000$ in this experiment. This improved accuracy comes at increased computational cost, so $TR$ provides a practical deployment trade-off. Moderate budgets, for example $TR=100$, yield clear gains at modest cost, while very large budgets, such as $TR=1000$, can approach the retrained reference but are more expensive.

\begin{table}
\centering
\caption{Retraining versus guidance under a structured denser-sensor deployment shift (interval-4 guidance). $\mathrm{GapClosed}$ is defined in equation~\eqref{eq:sec7_gapclosed} using time-averaged MAER, where the baseline is \texttt{cond\_ogi8} and the retrained reference is \texttt{cond\_ogi4\_retrain}. ``Cost factor'' reports the increase in the average total computation time over one full cycling run from $t=0$ to $t=24$ relative to the baseline sampler with $TR=20$.}
\label{tab:sec7_gapclosed}
\begin{tabular}{lcccc}
\hline
Case & Guidance grid & Reverse steps ($TR$) & Cost factor ($\times$) & $\mathrm{GapClosed}$ (\%) \\
\hline
\texttt{cond\_ogi4\_guid\_TR20} & interval 4 & 20   & 1.00 & 21.5 \\
\texttt{cond\_ogi4\_guid\_TR100} & interval 4 & 100  & 1.36 & 57.7 \\
\texttt{cond\_ogi4\_guid\_TR1000} & interval 4 & 1000 & 5.70 & 94.3 \\
\hline
\end{tabular}
\end{table}

\begin{figure}
  \centering
  \includegraphics[width=1.0\textwidth]{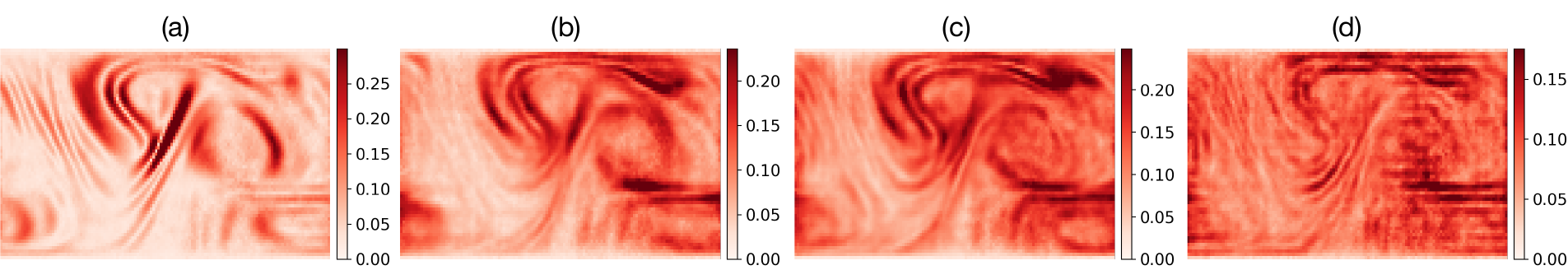}
  \caption{Ensemble spread fields under the structured interval-4 deployment setting. (a) \texttt{cond\_ogi4\_retrain}; (b) \texttt{cond\_ogi4\_guid\_TR20}; (c) \texttt{cond\_ogi4\_guid\_TR100}; (d) \texttt{cond\_ogi4\_guid\_TR1000}.}
\label{fig:guidance_uq_spreads}
\end{figure}

It is also useful to briefly examine how guidance affects uncertainty quantification. Although guidance is introduced here primarily as an inference-time accuracy correction, it acts by modifying the reverse-time score and can therefore change the ensemble distribution produced by the sampler. Since all ensembles in this section are generated with DDPM sampling, the spread reflects both model-induced uncertainty and stochasticity injected at each reverse step. Figure~\ref{fig:guidance_uq_spreads} shows the spread for the retrained interval-4 model \texttt{cond\_ogi4\_retrain} alongside the guided runs \texttt{cond\_ogi4\_guid\_TR20}, \texttt{cond\_ogi4\_guid\_TR100}, and \texttt{cond\_ogi4\_guid\_TR1000}. Compared with retraining, the guided cases with $TR=20$ and $TR=100$ retain relatively localized spread patterns and remain qualitatively similar in where uncertainty concentrates, whereas the $TR=1000$ guided case exhibits a smaller overall spread magnitude but a more spatially diffuse and speckled spread texture. This suggests that while increasing $TR$ can substantially improve the mean reconstruction under guidance, the resulting spread structure is not guaranteed to match that of a model retrained for the new sensor configuration.

This caveat is important when interpreting the random-layout results as well. In Figure~\ref{fig:obsguid_random}, all guided random-layout cases use $TR=1000$, because large reverse-step budgets are the most effective at adapting the pretrained model to strongly shifted sensor layouts. The price of this flexibility is twofold: first, the sampling cost is substantially higher than for the shorter guided chains used in the structured-layout experiments, and second, the associated uncertainty fields may become noisier and less localized, as suggested by the $TR=1000$ spread pattern in Figure~\ref{fig:guidance_uq_spreads}. A more systematic evaluation of guided uncertainty calibration, including spread, coverage, and rank statistics under guidance and their dependence on sampler stochasticity and step-dependent guidance schedules, is postponed to future work.

In this section, we showed that guidance can exploit deployment-time changes in the observation configuration without retraining DiffSRDA. In the structured interval-4 sensor-improvement scenario, guidance mitigates the degradation observed when the interval-8-trained model is evaluated under out-of-distribution conditioning \texttt{cond\_ogi4}, and the benefit increases with the reverse-step budget. In the more general random-layout experiments, guidance also improves reconstructions for fixed random sparse sensor configurations with both $N_s=128$ and $N_s=512$, showing that the approach is not restricted to regular-lattice densification alone. 
Using more reverse steps generally yields better reconstruction accuracy and can approach the retrained reference for sufficiently large budgets, but it also increases sampling cost and can alter the ensemble spread structure. In our testbed, a moderate budget around $TR\approx 100$ provides a practical compromise for the structured sensor-improvement setting, delivering clear accuracy gains at modest additional cost while retaining reasonably localized spread patterns. For more strongly shifted layouts, such as the fixed random sensor configurations considered here, guidance with very large reverse-step budgets can still be effective, but its computational costs and degradation in uncertainty spreads should be kept in mind.

While the goal of this section is mainly to illustrate how the score-based nature of diffusion sampling enables training-free corrections to the reverse trajectory, two practical caveats are worth noting. First, the guidance strength and blur scale require tuning. Overly aggressive updates (large $\kappa_\tau$) can overfit noisy observations and degrade reconstruction metrics even if the observation misfit improves, while excessive smoothing (large $\sigma_{\mathrm{blur}}$) can dilute informative local measurements. Step-dependent schedules for $\kappa_\tau$ are a natural extension for better balancing prior and likelihood influences across noise levels, but are beyond the scope of this work. Second, the efficient masked-residual updates in equations~\eqref{eq:sec7_clean_update}--\eqref{eq:sec7_clean_update_blur} rely on the mask-type on-grid observation operator used here. More general observation operators can be incorporated via the DPS-style gradient in Appendix~\ref{app:likelihood_guidance_derivation}, typically at higher computational cost and with additional stability considerations. In our experiments, the chosen settings ($\kappa_\tau=0.5$, $\sigma_{\mathrm{blur}}=1.5$) yielded stable improvements in both reconstruction error and observation consistency under the deployment shifts considered here.

\section{Discussion and concluding remarks}
\label{sec:conclusions}

A diffusion-based probabilistic super-resolution data assimilation framework, DiffSRDA, was developed to generate short HR analysis windows by combining low-cost LR forecast information with sparse noisy HR observations through conditional reverse diffusion sampling. The main conclusions of this study may be summarized as follows:
\begin{enumerate}
\item \textbf{DiffSRDA provides accurate and practical SRDA in a cycling setting.}  
Among the learned SRDA operators considered here, DiffSRDA consistently gives the best point-estimate accuracy and remains closest to the EnKF-HR baseline while avoiding the cost of repeatedly evolving HR forecast ensembles. A key practical result is that this base DiffSRDA analysis does not require long reverse chains in the present testbed: timestep respacing shows that only a small number of reverse steps is sufficient to retain essentially the same accuracy as full-chain sampling, making diffusion-based SRDA feasible for repeated forecast--analysis cycling.

\item \textbf{DiffSRDA provides useful ensemble-based uncertainty information.}  
Because DiffSRDA is generative, repeated conditional sampling directly yields an ensemble of HR analyses. The resulting spread fields capture physically meaningful uncertainty patterns and broadly agree with EnKF-HR in where uncertainty concentrates, especially in dynamically active regions with strong gradients and filamentary structure. Calibration diagnostics such as coverage and rank histograms indicate that the ensemble uncertainty is informative and stabilizes with a modest number of samples, although both DiffSRDA and EnKF-HR remain under-dispersed under the present pixelwise verification procedure.

\item \textbf{Deployment-time guidance provides flexible sensor adaptation, even with a simple and inexpensive guidance scheme.}  
Training-free observation-consistency guidance allows a fixed pretrained DiffSRDA model to adapt to changed sensor layouts at deployment, including both denser regular grids and fixed random sparse layouts, without retraining. Importantly, the objective of this work was not to optimize the strongest possible guidance method, but to demonstrate that even a computationally cheap stop-gradient guidance scheme can already make flexible sparse-sensor assimilation feasible in practice. One possible reason this lightweight guidance is effective is that DiffSRDA, being trained directly in an SRDA setting, may already learn a useful amortized black-box data-assimilation mapping from forecast and observation inputs to HR analyses. If so, deployment-time layout changes introduce only a partial distribution shift, and guidance need only act mainly as an observation-consistency correction rather than performing the full assimilation task from scratch. This is one of the potentially attractive features of SRDA-style frameworks. The results show that such simple guidance can recover a meaningful fraction of the retraining benefit and can also help absorb more general sensor-layout shifts beyond the regular grids used in training. At the same time, this flexibility should be distinguished from the base DiffSRDA result above: while short reverse chains are sufficient for accurate unguided DiffSRDA in the present setting, stronger guidance corrections may require substantially larger reverse-step budgets in order to accumulate smoothly. Those larger budgets increase sampling cost and can also degrade the qualitative structure of the ensemble spread if guidance is not tuned with uncertainty behavior in mind.
\end{enumerate}

One useful perspective on why DiffSRDA can be competitive with EnKF-HR is that the HR analysis is not inferred from observations alone. Rather, the conditional diffusion model learns a distribution over HR analysis windows given both the LR forecast window and the observations, so the LR forecast acts as a forecast-informed dynamical prior carrying information propagated from earlier cycles. As discussed in Appendix~\ref{sec:bayes_posterior_connection}, this is consistent with a forecast-informed Bayesian interpretation in which the learned conditional distribution reflects both observational information and dynamically propagated forecast context. In the present setting, this LR conditioning effectively carries forward part of the time-history information that would otherwise be supplied by repeatedly cycling an HR forecast model, which helps explain why the resulting analysis distribution is both accurate and physically plausible. By contrast, targeting the same posterior as a fully cycled HR-ensemble method would in general require the diffusion model itself to adapt to the cycling context \citep{hodyss2026using}, which would be considerably less attractive computationally. The results here suggest that conditioning on a reasonably faithful LR forecast can already recover much of that benefit within a single offline-trained generative model.

\paragraph{Limitations, related work, and future directions:}
One future direction that may be worthwhile to investigate is whether \emph{probabilistic surrogate forecast models}, capable of acting as standalone learned forecast models while remaining compatible with data assimilation, can offer a useful complementary alternative to SRDA when compared in terms of both accuracy and computational cost. In the present work, SR and DA are tightly coupled: DiffSRDA excels when low-cost forecast information and sparse observations are provided together, and the DA component helps regularize what would otherwise be a substantially more ill-posed SR-only prediction problem. Preliminary in-house experiments suggest that training and deploying the same framework as a pure super-resolution forecaster without built-in assimilation is much more challenging in the present chaotic multiscale setting, presumably because the temporal transition itself must then be learned more explicitly. This motivates future generative frameworks in which the dynamical transition is learned directly, so that fast HR probabilistic forecasts can be produced even without immediate observational input, and assimilation can then be carried out either through posterior sampling or through a separate filtering step. This broader direction is consistent with recent learned probabilistic forecasting systems such as GenCast \citep{price2023gencast}, with recent diffusion-based work that explicitly studies forecasting and assimilation together for PDE systems \citep{shysheya2024conditional}, and with stochastic-interpolant-based DA frameworks such as FlowDAS \citep{chen2025flowdas}, where learned transition dynamics and generative priors are treated within one framework.

A second important direction concerns how such generative forecast models should be coupled with DA. In the present work, posterior correction is implemented through a simple stop-gradient likelihood-guidance scheme. This was a deliberate choice: the goal was to show that flexible sparse-sensor assimilation is already possible with a lightweight and inexpensive mechanism, rather than to optimize the strongest available guidance strategy. More sophisticated posterior-sampling, score-correction, and ensemble-filtering approaches already exist \citep{rozet2023score,zhang2025improving,bao2024ensemble,bao2025nonlinear,zheng2024ensemble,garbuno2020interacting}, but they also introduce additional trade-offs, including higher computational cost, derivative or Jacobian requirements, stability issues, and possible changes in uncertainty behavior. The broader lesson is therefore problem-dependent: guidance is a viable tool for flexible sensor assimilation, but the most suitable guidance mechanism will depend on the dynamics, the observation operator, the computational budget, and the desired uncertainty properties.

More generally, an important methodological goal is to retain the deployment-time flexibility of guidance while reducing dependence on long reverse-time correction. In the present work, short reverse chains are sufficient for accurate base DiffSRDA, whereas stronger deployment-time corrections benefit from much larger reverse-step budgets. These observations suggest that future generative DA frameworks should aim to preserve guidance-like adaptability without sacrificing the efficiency of short-chain sampling. Several emerging alternatives may be relevant in this regard. Consistency models seek to replace iterative reverse diffusion with one- or few-step generation, either through direct conditional training or distillation from a pretrained diffusion model, and may therefore reduce inference cost or enable larger ensembles \citep{song2023consistency,song2023improved,hess2025fast}. Stochastic interpolants provide a unifying view of flows and diffusions that may offer new training and sampling strategies for conditional generation in dynamical systems \citep{albergo2023stochastic}, while Schr\"odinger-bridge and related transport-based viewpoints are attractive for DA because they emphasize distribution-to-distribution transport under observational constraints \citep{reich2019data,de2021diffusion,chen2024probabilistic,kim2025unified}. Exploring whether these frameworks can improve transition learning, calibration, data consistency, or deployment-time posterior correction in probabilistic SRDA is a promising direction for future work.
From the DA perspective itself, the likelihood-guidance mechanism developed here is tailored to a mask-type, on-grid observation operator with simple Gaussian assumptions. A natural next step is to extend the same inference-time prior-plus-likelihood idea to more general measurement processes, including off-grid sensors, indirect or integrated observations, partial state components, and more realistic observation-error models. Calibration also remains an open challenge, motivating improved likelihood models, schedule-aware or learned guidance strengths, and more explicit treatment of model error.

Another promising direction is to make these generative SRDA frameworks more interpretable and more useful for sensor design. Recent work such as Diff-SPORT suggests that explainable-AI tools, including Shapley-value-based attribution, can be combined with diffusion-based reconstruction to identify observation locations that matter most for accurate flow reconstruction and sensor placement \citep{vishwasrao2025diff,vinuesa2026explainable}. In future SRDA settings, such ideas may be especially useful for adaptive sensor placement, mobile-sensor steering, and physically interpretable observation-network design.

Finally, it will be important to test these ideas on higher-complexity flows and, eventually, on settings closer to operational forecasting and data assimilation, where both the dynamics and the observation process are considerably less idealized. In such settings, the most valuable outcome may not be a single preferred generative framework, but rather a family of probabilistic forecast-and-assimilation models that can flexibly span fast surrogate forecasting, posterior correction, and ensemble-based filtering, while offering varying degrees of physical interpretability depending on the needs of the application.

\section*{Acknowledgements}
This work was supported by the JSPS KAKENHI (Grant Number 25H00715) and Central Japan Railway Company. Some of the simulations and deep learning experiments were performed on the TSUBAME4.0 supercomputer provided by Institute of Science Tokyo and the Earth Simulator at the Japan Agency for Marine-Earth Science and Technology (JAMSTEC).

\bibliographystyle{abbrvnat} 
\bibliography{references.bib}

\appendix

\section{Interpreting DiffSRDA as a forecast-informed Bayesian posterior}
\label{sec:bayes_posterior_connection}

The DiffSRDA formulation admits a natural Bayesian interpretation. At each assimilation time, the analysis variable is a high-resolution (HR) state window $\mathbf{X}$ conditioned on two sources of information: sparse observations and low-resolution (LR) forecast information from a physics-based numerical model. Following \citet{hodyss2026using}, this may be viewed through an \emph{extended-likelihood} formulation, in which the forecast information enters as an additional likelihood factor, while the prior over HR windows remains fixed and climatological.

Let $\mathbf{y}$ denote the sparse HR observations available at the current assimilation time. Since the state variable is a window, the observation operator acts on the appropriate frame within $\mathbf{X}$, which in the present experiments is the current-time slice. The observation model is written as
\begin{equation}
\mathbf{y} = H\mathbf{X} + \boldsymbol{\eta},
\qquad
\boldsymbol{\eta} \sim \mathcal{N}(0,R_y),
\label{eq:obs_model_window}
\end{equation}
where $H$ is a sparse sampling operator on the HR grid, equivalently represented by a binary mask in the implementation, and $R_y$ is the observation-error covariance.

The forecast information is supplied by LR model outputs. The LR forecast sequence used as conditioning input is written as
\begin{equation}
\mathbf{f} \;\equiv\;
\bigl[\mathbf{x}_{t'-1}^{\mathrm{LR}},\mathbf{x}_{t'}^{\mathrm{LR}},\mathbf{x}_{t'+1}^{\mathrm{LR}}\bigr]
\;\in\;\mathbb{R}^{3n_{\mathrm{LR}}},
\label{eq:lr_forecast_def}
\end{equation}
where $t'$ is the LR time index aligned with the HR assimilation cycle. To relate the LR forecast sequence to the HR window $\mathbf{X}$, a linear operator $S$ is introduced that selects the corresponding HR frames within $\mathbf{X}$ and downsamples each selected frame to the LR grid. The LR forecasts are then modeled as a noisy proxy of the HR state through
\begin{equation}
\mathbf{f} = S\mathbf{X} + \boldsymbol{\xi},
\qquad
\boldsymbol{\xi} \sim \mathcal{N}(0,R_f),
\label{eq:forecast_model_window}
\end{equation}
where $R_f$ represents LR-model error together with representativeness error, that is, the mismatch between a downsampled HR field and the corresponding LR forecast. In this way, the super-resolution aspect of the problem is made explicit: the likelihood is posed in LR space, while the unknown remains HR.

Under the assumption that the observation and forecast errors are conditionally independent given $\mathbf{X}$, the resulting forecast-informed posterior takes the form
\begin{equation}
p(\mathbf{X}\mid \mathbf{f},\mathbf{y})
\;\propto\;
p(\mathbf{y}\mid \mathbf{X})\,p(\mathbf{f}\mid \mathbf{X})\,p_{\mathrm{clim}}(\mathbf{X}),
\label{eq:extlik_posterior_window}
\end{equation}
where $p_{\mathrm{clim}}(\mathbf{X})$ denotes a climatological prior over HR windows, represented in practice by the HR training distribution.

DiffSRDA is trained as a conditional diffusion model on HR windows $\mathbf{X}_0$ with conditioning input $\mathbf{c}=(\mathbf{f},\mathbf{y})$. Under forward noising,
\begin{equation}
\mathbf{X}_\tau
=
\sqrt{\bar{\alpha}_\tau}\,\mathbf{X}_0
+
\sqrt{1-\bar{\alpha}_\tau}\,\boldsymbol{\varepsilon},
\qquad
\boldsymbol{\varepsilon}\sim\mathcal{N}(0,I),
\label{eq:forward_window}
\end{equation}
the denoiser predicts $\boldsymbol{\varepsilon}_\theta(\mathbf{X}_\tau,\tau,\mathbf{c})$ from $(\mathbf{X}_\tau,\tau,\mathbf{c})$. An estimator of the conditional score is thereby obtained as
\begin{equation}
\nabla_{\mathbf{X}_\tau}\log p(\mathbf{X}_\tau\mid \mathbf{c})
\;\approx\;
-\frac{1}{\sqrt{1-\bar{\alpha}_\tau}}\,
\boldsymbol{\varepsilon}_\theta(\mathbf{X}_\tau,\tau,\mathbf{c}),
\label{eq:score_from_eps_window}
\end{equation}
together with the associated clean-state estimate given by Tweedie's formula,
\begin{equation}
\widehat{\mathbf{X}}_0(\mathbf{X}_\tau,\mathbf{c})
=
\frac{1}{\sqrt{\bar{\alpha}_\tau}}
\left(
\mathbf{X}_\tau - \sqrt{1-\bar{\alpha}_\tau}\,\boldsymbol{\varepsilon}_\theta(\mathbf{X}_\tau,\tau,\mathbf{c})
\right)
\;\approx\;
\mathbb{E}[\mathbf{X}_0\mid \mathbf{X}_\tau,\mathbf{c}].
\label{eq:tweedie_window}
\end{equation}
Reverse-time sampling approximately integrates this conditional score field and yields samples whose terminal distribution approximates the learned conditional distribution $p(\mathbf{X}\mid \mathbf{c}) = p(\mathbf{X}\mid \mathbf{f},\mathbf{y})$. In this sense, DiffSRDA may be interpreted as learning an approximation to the forecast-informed posterior in equation \eqref{eq:extlik_posterior_window}, rather than explicitly constructing that posterior from a prescribed analytical prior and likelihood.

In the special case where $p_{\mathrm{clim}}(\mathbf{X})$, $p(\mathbf{y}\mid \mathbf{X})$, and $p(\mathbf{f}\mid \mathbf{X})$ are all linear--Gaussian, the posterior in equation \eqref{eq:extlik_posterior_window} is itself Gaussian. In that setting, diffusion-based sampling is consistent with this posterior, and the relationship between the posterior mean and covariance and the reverse-time diffusion dynamics can be analyzed explicitly, as shown by \citet{hodyss2026using}. Outside this special case, however, the value of the present interpretation is mainly conceptual: it clarifies that DiffSRDA is not merely learning a black-box LR-to-HR super-resolution map, but rather a conditional distribution over HR analysis windows informed jointly by observations and forecast information.

This viewpoint is useful for interpreting the results in the main text. The LR forecast conditioning does not provide only static contextual input; dynamical information carried forward from previous forecast--analysis cycles is also conveyed through it. As a result, the conditional distribution learned by DiffSRDA acquires a filtering-like character: current observations are combined with a forecast-informed dynamical context, even though the posterior is represented implicitly through a learned conditional score rather than through an explicit Kalman-style update. This helps explain why DiffSRDA can approach the accuracy of sequential DA baselines while still operating as a learned generative SRDA framework.

\section{Derivation of observation-consistency guidance expression}
\label{app:likelihood_guidance_derivation}

This appendix derives the training-free guidance used to incorporate additional inference-time observations during reverse diffusion sampling. The trained conditional diffusion model is assumed to represent a conditional distribution $p_\theta(\mathbf{X}_0\mid \mathbf{c})$ over the clean HR analysis window $\mathbf{X}_0\in\mathbb{R}^{5n_{\mathrm{HR}}}$ given conditioning input $\mathbf{c}$. At inference time, additional observations $\mathbf{y}$ are incorporated through a Gaussian likelihood, and an approximate guided posterior is constructed for use during the reverse diffusion process.

The guided target density over the clean state is taken as
\begin{equation}
\tilde p(\mathbf{X}_0\mid \mathbf{c},\mathbf{y})
\propto
p_\theta(\mathbf{X}_0\mid \mathbf{c})\,p(\mathbf{y}\mid \mathbf{X}_0),
\label{eq:app_guided_posterior}
\end{equation}
where, throughout this work, a linear observation model with additive Gaussian noise is assumed:
\begin{equation}
\mathbf{y} = H\mathbf{X}_0 + \boldsymbol{\eta},
\qquad
\boldsymbol{\eta}\sim\mathcal{N}(0,R).
\label{eq:app_obs_model}
\end{equation}
Here $H$ is understood to include the appropriate slicing of the HR analysis window together with the sparse on-grid sampling operator. The resulting log-likelihood is
\begin{equation}
\log p(\mathbf{y}\mid \mathbf{X}_0)
=
-\frac12 (\mathbf{y}-H\mathbf{X}_0)^\top R^{-1}(\mathbf{y}-H\mathbf{X}_0)
+ \mathrm{const},
\label{eq:app_loglik}
\end{equation}
and its gradient with respect to the clean state is
\begin{equation}
\nabla_{\mathbf{X}_0}\log p(\mathbf{y}\mid \mathbf{X}_0)
=
H^\top R^{-1}(\mathbf{y}-H\mathbf{X}_0).
\label{eq:app_grad_loglik}
\end{equation}

Reverse diffusion sampling evolves a noisy variable $\mathbf{X}_\tau$ rather than $\mathbf{X}_0$, so the diffusion-time posterior score
$\nabla_{\mathbf{X}_\tau}\log \tilde p_\tau(\mathbf{X}_\tau\mid \mathbf{c},\mathbf{y})$
is required. The clean-space guided target in equation \eqref{eq:app_guided_posterior} is therefore translated into diffusion-time marginals induced by the forward diffusion kernel. Let $q_\tau(\mathbf{X}_\tau\mid \mathbf{X}_0)$ denote the forward kernel at diffusion step $\tau$,
\begin{equation}
q_\tau(\mathbf{X}_\tau\mid \mathbf{X}_0)
=
\mathcal{N}\!\left(
\mathbf{X}_\tau;\sqrt{\bar\alpha_\tau}\,\mathbf{X}_0,\,
(1-\bar\alpha_\tau)I
\right),
\label{eq:app_forward_kernel}
\end{equation}
and let the corresponding diffusion-time marginals be defined by
\begin{equation}
p_{\theta,\tau}(\mathbf{X}_\tau\mid \mathbf{c})
=
\int q_\tau(\mathbf{X}_\tau\mid \mathbf{X}_0)\,p_\theta(\mathbf{X}_0\mid \mathbf{c})\,d\mathbf{X}_0,
\qquad
\tilde p_\tau(\mathbf{X}_\tau\mid \mathbf{c},\mathbf{y})
=
\int q_\tau(\mathbf{X}_\tau\mid \mathbf{X}_0)\,\tilde p(\mathbf{X}_0\mid \mathbf{c},\mathbf{y})\,d\mathbf{X}_0.
\label{eq:app_marginals}
\end{equation}
Substituting equation \eqref{eq:app_guided_posterior} into equation \eqref{eq:app_marginals} gives
\begin{equation}
\tilde p_\tau(\mathbf{X}_\tau\mid \mathbf{c},\mathbf{y})
\propto
\int q_\tau(\mathbf{X}_\tau\mid \mathbf{X}_0)\,
p_\theta(\mathbf{X}_0\mid \mathbf{c})\,
p(\mathbf{y}\mid \mathbf{X}_0)\,
d\mathbf{X}_0.
\label{eq:app_tilde_marginal_expand}
\end{equation}

To separate this guided marginal into an unguided term in $\mathbf{X}_\tau$ and an observation-likelihood term, Bayes' rule is applied under the induced joint density
$p_\theta(\mathbf{X}_0\mid \mathbf{c})\,q_\tau(\mathbf{X}_\tau\mid \mathbf{X}_0)$, leading to
\begin{equation}
p_\theta(\mathbf{X}_0\mid \mathbf{X}_\tau,\mathbf{c})
:=
\frac{
q_\tau(\mathbf{X}_\tau\mid \mathbf{X}_0)\,p_\theta(\mathbf{X}_0\mid \mathbf{c})
}{
p_{\theta,\tau}(\mathbf{X}_\tau\mid \mathbf{c})
}.
\label{eq:app_posterior_x0_given_xt}
\end{equation}
Equivalently,
\begin{equation}
q_\tau(\mathbf{X}_\tau\mid \mathbf{X}_0)\,p_\theta(\mathbf{X}_0\mid \mathbf{c})
=
p_{\theta,\tau}(\mathbf{X}_\tau\mid \mathbf{c})\,
p_\theta(\mathbf{X}_0\mid \mathbf{X}_\tau,\mathbf{c}).
\label{eq:app_joint_factor_identity}
\end{equation}
Substitution of equation \eqref{eq:app_joint_factor_identity} into equation \eqref{eq:app_tilde_marginal_expand} yields the exact factorization
\begin{equation}
\tilde p_\tau(\mathbf{X}_\tau\mid \mathbf{c},\mathbf{y})
\propto
p_{\theta,\tau}(\mathbf{X}_\tau\mid \mathbf{c})\,
\underbrace{
\int p(\mathbf{y}\mid \mathbf{X}_0)\,
p_\theta(\mathbf{X}_0\mid \mathbf{X}_\tau,\mathbf{c})\,
d\mathbf{X}_0
}_{=:~p_{\theta,\tau}(\mathbf{y}\mid \mathbf{X}_\tau,\mathbf{c})},
\label{eq:app_factorization}
\end{equation}
where
\begin{equation}
p_{\theta,\tau}(\mathbf{y}\mid \mathbf{X}_\tau,\mathbf{c})
:=
\int p(\mathbf{y}\mid \mathbf{X}_0)\,
p_\theta(\mathbf{X}_0\mid \mathbf{X}_\tau,\mathbf{c})\,
d\mathbf{X}_0
\label{eq:app_marginal_likelihood}
\end{equation}
is the model-induced diffusion-time conditional likelihood of $\mathbf{y}$ given $\mathbf{X}_\tau$. Taking the gradient of the log of equation \eqref{eq:app_factorization} gives the exact diffusion-time score decomposition
\begin{equation}
\nabla_{\mathbf{X}_\tau}\log \tilde p_\tau(\mathbf{X}_\tau\mid \mathbf{c},\mathbf{y})
=
\nabla_{\mathbf{X}_\tau}\log p_{\theta,\tau}(\mathbf{X}_\tau\mid \mathbf{c})
+
\nabla_{\mathbf{X}_\tau}\log p_{\theta,\tau}(\mathbf{y}\mid \mathbf{X}_\tau,\mathbf{c}),
\label{eq:app_score_decomp_exact}
\end{equation}
where the proportionality constant in equation \eqref{eq:app_factorization} depends only on $(\mathbf{c},\mathbf{y})$ and therefore contributes no gradient with respect to $\mathbf{X}_\tau$. The first term is the unguided model score, while the second term is the diffusion-time measurement score. In general, the measurement term is intractable because $p_{\theta,\tau}(\mathbf{y}\mid \mathbf{X}_\tau,\mathbf{c})$ is defined by the marginalization in equation \eqref{eq:app_marginal_likelihood}, that is, by averaging the clean-space likelihood over the conditional distribution $p_\theta(\mathbf{X}_0\mid \mathbf{X}_\tau,\mathbf{c})$.

A standard plug-in approximation, used in diffusion posterior sampling (DPS) and related guided samplers, replaces this intractable diffusion-time measurement score by evaluating the clean-space likelihood at a denoised estimate of $\mathbf{X}_0$ computed from the current iterate $\mathbf{X}_\tau$ \citep{chung2022diffusion}. Concretely, the approximation
\begin{equation}
\nabla_{\mathbf{X}_\tau}\log p_{\theta,\tau}(\mathbf{y}\mid \mathbf{X}_\tau,\mathbf{c})
\approx
\nabla_{\mathbf{X}_\tau}\log p\!\left(\mathbf{y}\mid \hat{\mathbf{X}}_0(\mathbf{X}_\tau,\tau,\mathbf{c})\right)
\label{eq:app_dps_plugin}
\end{equation}
is used, where $\hat{\mathbf{X}}_0(\mathbf{X}_\tau,\tau,\mathbf{c})$ is the clean-state estimate implied by the denoiser output at diffusion step $\tau$. Under the forward relation
\begin{equation}
\mathbf{X}_\tau
=
\sqrt{\bar\alpha_\tau}\,\mathbf{X}_0
+
\sqrt{1-\bar\alpha_\tau}\,\boldsymbol{\varepsilon},
\qquad
\boldsymbol{\varepsilon}\sim\mathcal{N}(0,I),
\label{eq:app_forward_relation}
\end{equation}
the $\boldsymbol{\varepsilon}$-parameterized denoiser yields the standard clean estimate
\begin{equation}
\hat{\mathbf{X}}_0(\mathbf{X}_\tau,\tau,\mathbf{c})
=
\frac{
\mathbf{X}_\tau-\sqrt{1-\bar\alpha_\tau}\,\boldsymbol{\varepsilon}_\theta(\mathbf{X}_\tau,\tau,\mathbf{c})
}{
\sqrt{\bar\alpha_\tau}
}.
\label{eq:app_x0hat}
\end{equation}
The corresponding approximation of the model score is
\begin{equation}
s_\theta(\mathbf{X}_\tau,\tau,\mathbf{c})
:=
\nabla_{\mathbf{X}_\tau}\log p_{\theta,\tau}(\mathbf{X}_\tau\mid \mathbf{c})
\approx
-\frac{1}{\sqrt{1-\bar\alpha_\tau}}\,
\boldsymbol{\varepsilon}_\theta(\mathbf{X}_\tau,\tau,\mathbf{c}),
\label{eq:app_score_from_eps}
\end{equation}
and equation \eqref{eq:app_x0hat} can equivalently be written in Tweedie form as
\begin{equation}
\hat{\mathbf{X}}_0(\mathbf{X}_\tau,\tau,\mathbf{c})
=
\frac{1}{\sqrt{\bar\alpha_\tau}}
\left(
\mathbf{X}_\tau+(1-\bar\alpha_\tau)\,s_\theta(\mathbf{X}_\tau,\tau,\mathbf{c})
\right).
\label{eq:app_tweedie}
\end{equation}

The approximate measurement gradient in diffusion space is then derived under equation \eqref{eq:app_dps_plugin}. Define
\begin{equation}
g_\tau
:=
\nabla_{\mathbf{X}_\tau}\log p\!\left(\mathbf{y}\mid \hat{\mathbf{X}}_0(\mathbf{X}_\tau,\tau,\mathbf{c})\right).
\label{eq:app_g_def}
\end{equation}
By the chain rule,
\begin{equation}
g_\tau
=
\left(\frac{\partial \hat{\mathbf{X}}_0}{\partial \mathbf{X}_\tau}\right)^{\!\top}
\nabla_{\mathbf{X}_0}\log p(\mathbf{y}\mid \mathbf{X}_0)\Big|_{\mathbf{X}_0=\hat{\mathbf{X}}_0}.
\label{eq:app_chain_rule}
\end{equation}
For completeness, if the observation operator is a possibly nonlinear forward map $A(\cdot)$ with Gaussian noise,
$\mathbf{y}=A(\mathbf{X}_0)+\boldsymbol{\eta}$,
$\boldsymbol{\eta}\sim\mathcal{N}(0,R)$,
then
\begin{equation}
\nabla_{\mathbf{X}_0}\log p(\mathbf{y}\mid \mathbf{X}_0)
=
J_A(\mathbf{X}_0)^\top R^{-1}\!\left(\mathbf{y}-A(\mathbf{X}_0)\right),
\label{eq:app_grad_loglik_general}
\end{equation}
where $J_A(\mathbf{X}_0):=\nabla_{\mathbf{X}_0}A(\mathbf{X}_0)$ is the Jacobian of the forward operator. Substitution of equation \eqref{eq:app_grad_loglik_general} into equation \eqref{eq:app_chain_rule} gives
\begin{equation}
g_\tau
=
\left(\frac{\partial \hat{\mathbf{X}}_0}{\partial \mathbf{X}_\tau}\right)^{\!\top}
J_A(\hat{\mathbf{X}}_0)^\top R^{-1}\!\left(\mathbf{y}-A(\hat{\mathbf{X}}_0)\right).
\label{eq:app_g_exact_general}
\end{equation}
In the present work, $A(\mathbf{X}_0)=H\mathbf{X}_0$ is linear, so $J_A(\cdot)=H$, and equation \eqref{eq:app_g_exact_general} reduces to
\begin{equation}
g_\tau
=
\left(\frac{\partial \hat{\mathbf{X}}_0}{\partial \mathbf{X}_\tau}\right)^{\!\top}
H^\top R^{-1}\!\left(\mathbf{y}-H\hat{\mathbf{X}}_0\right).
\label{eq:app_g_exact}
\end{equation}

Full DPS evaluates equation \eqref{eq:app_g_exact} by differentiating through the denoiser, which requires Jacobian--vector products involving
$\partial \boldsymbol{\varepsilon}_\theta/\partial \mathbf{X}_\tau$
at every reverse step \citep{chung2022diffusion}. To avoid this backward pass, a stop-gradient (straight-through) Jacobian approximation is adopted based on equation \eqref{eq:app_x0hat}. Differentiation of equation \eqref{eq:app_x0hat} gives
\begin{equation}
\frac{\partial \hat{\mathbf{X}}_0}{\partial \mathbf{X}_\tau}
=
\frac{1}{\sqrt{\bar\alpha_\tau}}I
-
\frac{\sqrt{1-\bar\alpha_\tau}}{\sqrt{\bar\alpha_\tau}}\,
\frac{\partial \boldsymbol{\varepsilon}_\theta}{\partial \mathbf{X}_\tau},
\label{eq:app_jac_exact}
\end{equation}
and this is approximated by retaining only the explicit linear scaling term,
\begin{equation}
\frac{\partial \hat{\mathbf{X}}_0}{\partial \mathbf{X}_\tau}
\approx
\frac{1}{\sqrt{\bar\alpha_\tau}}\,I.
\label{eq:app_jac_approx}
\end{equation}
Substitution of equation \eqref{eq:app_jac_approx} into equation \eqref{eq:app_g_exact} yields the inexpensive likelihood-gradient approximation
\begin{equation}
g_\tau
\approx
\frac{1}{\sqrt{\bar\alpha_\tau}}\,
H^\top R^{-1}\!\left(\mathbf{y}-H\hat{\mathbf{X}}_0\right).
\label{eq:app_g_approx}
\end{equation}
For isotropic observation noise $R=\sigma_{\mathrm{obs}}^2 I$, equation \eqref{eq:app_g_approx} becomes
\begin{equation}
g_\tau
\approx
\frac{1}{\sigma_{\mathrm{obs}}^2\sqrt{\bar\alpha_\tau}}\,
H^\top\!\left(\mathbf{y}-H\hat{\mathbf{X}}_0\right).
\label{eq:app_g_approx_iso}
\end{equation}

The score decomposition in equation \eqref{eq:app_score_decomp_exact} is implemented by defining a guided score
\begin{equation}
s_{\mathrm{post}}(\mathbf{X}_\tau,\tau,\mathbf{c},\mathbf{y})
:=
s_\theta(\mathbf{X}_\tau,\tau,\mathbf{c}) + \lambda_\tau\, g_\tau,
\label{eq:app_score_post}
\end{equation}
where $\lambda_\tau$ controls the guidance strength. Under the $\boldsymbol{\varepsilon}$-parameterization, adding $\lambda_\tau g_\tau$ is equivalent to modifying the noise prediction, since
$s_\theta=-\boldsymbol{\varepsilon}_\theta/\sqrt{1-\bar\alpha_\tau}$.
Thus,
\begin{equation}
\boldsymbol{\varepsilon}_{\mathrm{post}}
=
\boldsymbol{\varepsilon}_\theta
-
\sqrt{1-\bar\alpha_\tau}\,\lambda_\tau\, g_\tau.
\label{eq:app_eps_post}
\end{equation}
The same guidance can also be expressed as a direct correction of the clean estimate. Using equation \eqref{eq:app_tweedie} with $s_{\mathrm{post}}$ and holding $\mathbf{X}_\tau$ fixed, the induced change in the clean estimate is
\begin{equation}
\Delta \hat{\mathbf{X}}_0
=
\frac{1-\bar\alpha_\tau}{\sqrt{\bar\alpha_\tau}}\,
\lambda_\tau\, g_\tau.
\label{eq:app_delta_x0_general}
\end{equation}
Substitution of equation \eqref{eq:app_g_approx} yields a schedule-aware residual correction:
\begin{equation}
\Delta \hat{\mathbf{X}}_0
\approx
\lambda_\tau\,
\frac{1-\bar\alpha_\tau}{\bar\alpha_\tau}\,
H^\top R^{-1}\!\left(\mathbf{y}-H\hat{\mathbf{X}}_0\right),
\label{eq:app_delta_x0_general_R}
\end{equation}
which increases $\log p(\mathbf{y}\mid \mathbf{X}_0)$ and reduces the misfit $\mathbf{y}-H\hat{\mathbf{X}}_0$.

In the present experiments, $H$ is a sparse on-grid selection operator acting on the HR analysis window. Define
\begin{equation}
M:=H^\top H,
\label{eq:app_M_def}
\end{equation}
which is diagonal with binary entries for pure selection. If the observations are embedded into the full tensor space as
$\tilde{\mathbf{y}}:=H^\top \mathbf{y}$,
then
\begin{equation}
H^\top\!\left(\mathbf{y}-H\hat{\mathbf{X}}_0\right)
=
\tilde{\mathbf{y}} - M\hat{\mathbf{X}}_0
=
M\odot(\tilde{\mathbf{y}}-\hat{\mathbf{X}}_0).
\label{eq:app_mask_residual}
\end{equation}
Guidance is therefore implemented as the clean-space correction
\begin{equation}
\hat{\mathbf{X}}_0
\leftarrow
\hat{\mathbf{X}}_0 + \kappa_\tau\,\bigl(M\odot(\tilde{\mathbf{y}}-\hat{\mathbf{X}}_0)\bigr),
\label{eq:app_clean_update}
\end{equation}
where $\kappa_\tau$ is a tunable step size that absorbs the schedule and likelihood scaling in equation \eqref{eq:app_delta_x0_general_R}. In particular, $\kappa_\tau$ is proportional to
$\lambda_\tau(1-\bar\alpha_\tau)/\bar\alpha_\tau$
and may also absorb factors from $R^{-1}$ when $R$ is diagonal. A soft-guidance variant is also considered, in which the lifted residual is spatially spread by a channel-wise Gaussian blur operator $\mathcal{G}_{\sigma_{\mathrm{blur}}}$:
\begin{equation}
\hat{\mathbf{X}}_0
\leftarrow
\hat{\mathbf{X}}_0
+
\kappa_\tau\,
\mathcal{G}_{\sigma_{\mathrm{blur}}}\!\bigl(M\odot(\tilde{\mathbf{y}}-\hat{\mathbf{X}}_0)\bigr).
\label{eq:app_clean_update_blur}
\end{equation}
This blur acts as a heuristic preconditioner and is not the exact gradient of the Gaussian likelihood.

After modification of $\hat{\mathbf{X}}_0$, a noise estimate consistent with the forward relation in equation \eqref{eq:app_forward_relation} may optionally be recomputed:
\begin{equation}
\boldsymbol{\varepsilon}_{\mathrm{rec}}
=
\frac{
\mathbf{X}_\tau-\sqrt{\bar\alpha_\tau}\,\hat{\mathbf{X}}_0
}{
\sqrt{1-\bar\alpha_\tau}
}.
\label{eq:app_eps_recompute}
\end{equation}
This keeps $(\hat{\mathbf{X}}_0,\boldsymbol{\varepsilon}_{\mathrm{rec}})$ consistent with the current $\mathbf{X}_\tau$ after the guidance edit.

Finally, let $\{\tau_k\}_{k=0}^{K}$ be the possibly respaced reverse-time index set with $\tau_K=T$ and $\tau_0=0$. At step $k$, $\hat{\mathbf{X}}_0$ is computed from equation \eqref{eq:app_x0hat}, equation \eqref{eq:app_clean_update} or equation \eqref{eq:app_clean_update_blur} is applied, $\boldsymbol{\varepsilon}_{\mathrm{rec}}$ may optionally be computed from equation \eqref{eq:app_eps_recompute}, and the reverse chain is then advanced from $\mathbf{X}_{\tau_k}$ to $\mathbf{X}_{\tau_{k-1}}$. A unified DDPM/DDIM-style update is
\begin{equation}
\mathbf{X}_{\tau_{k-1}}
=
\sqrt{\bar\alpha_{\tau_{k-1}}}\,\hat{\mathbf{X}}_0
+
\sqrt{1-\bar\alpha_{\tau_{k-1}}-\sigma_{\tau_k}^2}\;\boldsymbol{\varepsilon}_{\mathrm{rec}}
+
\sigma_{\tau_k}\,\mathbf{z},
\qquad
\mathbf{z}\sim\mathcal{N}(0,I),
\label{eq:app_reverse_unified}
\end{equation}
where $\sigma_{\tau_k}\ge 0$ is the per-step noise level determined by the chosen sampler, with DDIM corresponding to $\sigma_{\tau_k}=0$. In this formulation, guidance influences the reverse dynamics through the corrected clean estimate $\hat{\mathbf{X}}_0$, and through $\boldsymbol{\varepsilon}_{\mathrm{rec}}$ if it is recomputed.

It is sometimes useful to interpret the likelihood-score approximation in equation \eqref{eq:app_g_approx} through the generic lifted-residual template summarized in diffusion inverse-problem surveys \citep{daras2024survey}. In that notation, many explicit measurement-score approximations can be written schematically as
\begin{equation}
\nabla_{\mathbf{X}_\tau}\log p_\tau(\mathbf{y}\mid \mathbf{X}_\tau,\mathbf{c})
\approx
-\frac{L_\tau\,M_\tau}{G_\tau},
\label{eq:app_daras_template}
\end{equation}
where $M_\tau$ denotes a measurement misfit, $L_\tau$ lifts the misfit back to the state space, and $G_\tau$ sets the overall strength, often through likelihood and diffusion-time scalings.

For a generic forward operator $A(\cdot)$ with Gaussian noise,
$\mathbf{y}=A(\mathbf{X}_0)+\boldsymbol{\eta}$,
$\boldsymbol{\eta}\sim\mathcal{N}(0,R)$,
the DPS-style approximation in equation \eqref{eq:app_g_exact_general}, combined with the Jacobian simplification in equation \eqref{eq:app_jac_approx}, yields
\begin{equation}
\nabla_{\mathbf{X}_\tau}\log p_\tau(\mathbf{y}\mid \mathbf{X}_\tau,\mathbf{c})
\approx
\frac{1}{\sqrt{\bar\alpha_\tau}}\,
J_A(\hat{\mathbf{X}}_0)^\top R^{-1}\!\left(\mathbf{y}-A(\hat{\mathbf{X}}_0)\right).
\label{eq:app_daras_lifted_residual}
\end{equation}
This corresponds to choosing
$M_\tau := \mathbf{y}-A(\hat{\mathbf{X}}_0)$,
$L_\tau := -\bar\alpha_\tau^{-1/2}J_A(\hat{\mathbf{X}}_0)^\top$,
and $G_\tau := R$
in equation \eqref{eq:app_daras_template}. In the linear case $A(\mathbf{X}_0)=H\mathbf{X}_0$, one has $J_A(\cdot)=H$, and equation \eqref{eq:app_daras_lifted_residual} reduces to equation \eqref{eq:app_g_approx}.

It is emphasized that equation \eqref{eq:app_jac_approx} is a stop-gradient approximation motivated by efficiency and by the simplicity of the sparse on-grid masking operator used here; in general, it does not yield an exact diffusion-time posterior score. For more challenging cases, including nonlinear observation operators or stronger guidance, covariance-aware likelihood corrections may improve stability \citep{rozet2023score}.

\section{SRDA using latent diffusion as a forward-looking scalability ablation}
\label{app:SRDA_ldm}

As a forward-looking scalability ablation, latent diffusion is also evaluated for SRDA. The motivation is that the cost of reverse diffusion grows with both the number of reverse steps and the state dimension, so diffusion in a compressed latent space may become attractive in future settings with much larger HR state dimensions. The purpose of this appendix is therefore not to present LatDiffSRDA as a competing mainline method, but rather to assess whether compressed-space diffusion could offer a practical route toward lower-cost SRDA in more demanding applications.

\begin{figure}
  \centering
  \includegraphics[width=1.0\textwidth]{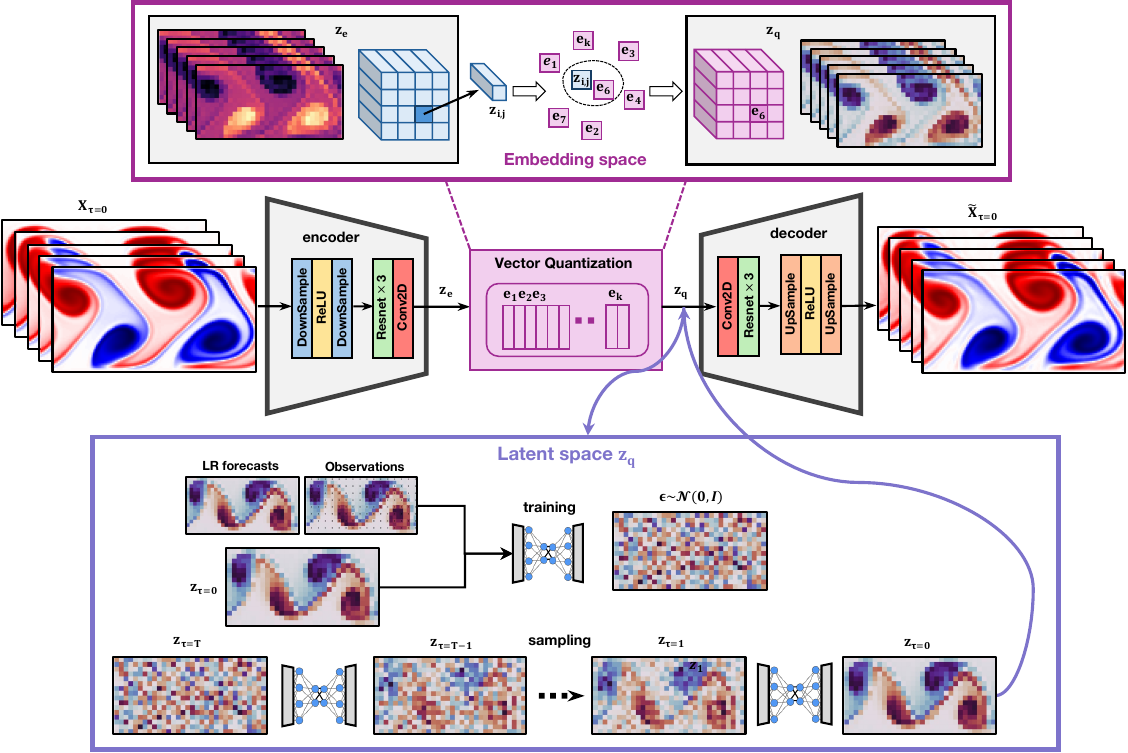}
  \caption{Conditional latent diffusion model used for LatDiffSRDA. The middle block shows the vector-quantized variational autoencoder (VQ-VAE): an HR input window $\mathbf{X}_{\tau=0}$ is encoded to continuous latents $\mathbf{z}_e$, mapped to vector-quantized latents $\mathbf{z}_q$ in the codebook embedding space, and decoded to a reconstruction $\tilde{\mathbf{X}}_{\tau=0}$. The top block illustrates the vector quantization step, where each local latent vector is replaced by its nearest embedding. The bottom block shows conditional diffusion operating in latent space: the denoiser is trained and sampled on $\mathbf{z}_q$, with LR forecasts and observations provided as conditioning inputs.}
  \label{fig:LDM_vqvae_illustration}
\end{figure}

\begin{figure}
  \centering
  \includegraphics[width=1.0\textwidth]{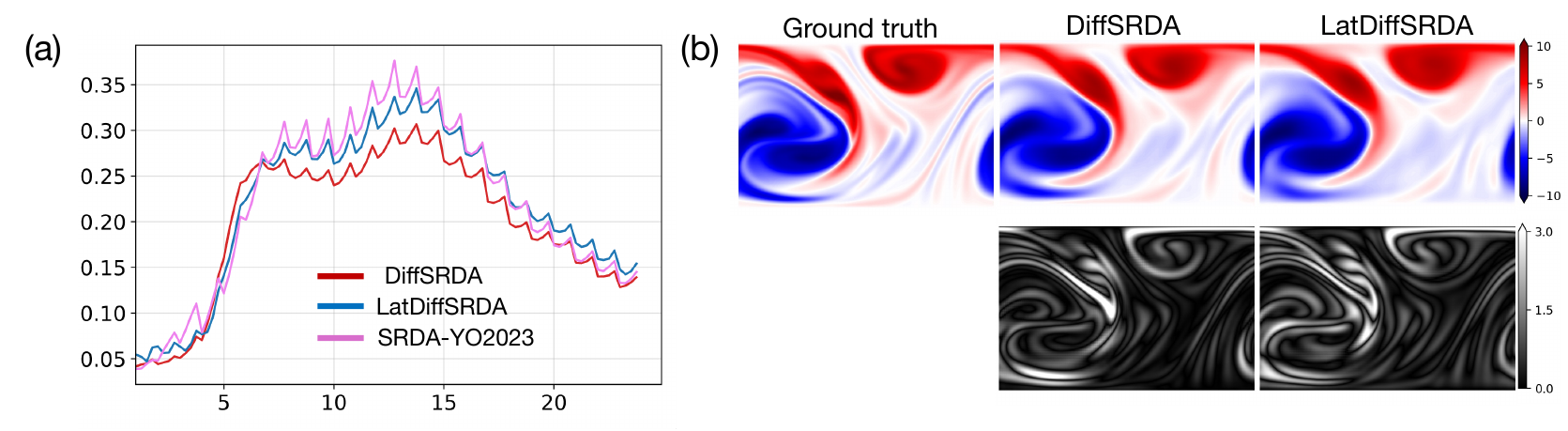}
  \caption{Forward-looking scalability ablation comparing pixel-space DiffSRDA and latent LatDiffSRDA. (a) MAER of the vorticity field versus time, averaged over the full test set, with SRDA-YO2023 shown for reference. (b) Representative vorticity snapshot at $t=14$ and the corresponding absolute error relative to the UHR reference for DiffSRDA and LatDiffSRDA. Five reverse steps are used for both methods.}
  \label{fig:ldm_ablation}
\end{figure}

LatDiffSRDA follows the same SRDA cycling protocol as DiffSRDA. Low-resolution (LR) forecast information together with sparse observations is used as conditioning input, and the SRDA operator outputs a high-resolution (HR) analysis window that is used to restart the next LR forecast cycle. The only change is the space in which diffusion is performed. Instead of running diffusion directly on HR windows in physical space, LatDiffSRDA performs diffusion in the latent space of a separately trained vector-quantized variational autoencoder (VQ-VAE). After VQ-VAE training, an HR window is mapped by the encoder to a compact latent tensor, quantized in the codebook embedding space, and decoded back to physical space. The VQ-VAE is then kept fixed, while the conditional diffusion model is trained and sampled in the quantized latent space. Figure~\ref{fig:LDM_vqvae_illustration} illustrates this pipeline. Architectural details, including the VQ-VAE and latent tensor shapes, are provided in Appendix~\ref{app:architectures}.

Figure~\ref{fig:ldm_ablation} compares LatDiffSRDA and DiffSRDA under the same evaluation protocol and a fixed reverse-step budget. In the present moderate-size testbed, LatDiffSRDA is broadly comparable to DiffSRDA but does not provide a clear accuracy advantage. A plausible explanation is that the VQ-VAE introduces an additional approximation through compression and quantization, which can degrade small-scale fidelity and affect the representation of uncertainty. For this reason, the latent formulation is not adopted as the mainline method in this study.

On the other hand, LatDiffSRDA can reduce sampling cost. For the full reverse chain with $T=1000$ reverse steps, LatDiffSRDA required approximately $362$ seconds of wall-clock time per full SRDA cycle from $t=0$ to $t=24$, compared with approximately $562$ seconds for pixel-space DiffSRDA under the same setting. In the present testbed, this reduction in cost does not outweigh the lack of a clear accuracy benefit. However, the balance may shift in future applications where the HR state dimension is much larger and pixel-space diffusion becomes substantially more expensive.

Two follow-on directions are worth noting. First, it may be beneficial to learn smoother or more task-aware latent representations, beyond reconstruction-oriented VQ-VAE training, so that the compressed space better preserves the flow structures most relevant for analysis and uncertainty. More broadly, recent work on machine learning for dynamical systems highlights the importance of informative low-dimensional representations and reduced-order latent spaces for modeling complex spatiotemporal dynamics \citep{cheng2023machine}. Within that broader perspective, promising directions include conditional neural-field latent diffusion models for spatiotemporal turbulence, which offer advantages such as geometry flexibility, mesh-agnostic representation, and compact generation of extended spatiotemporal fields in latent space \citep{du2024conditional}, as well as $\beta$-variational-autoencoder-based reduced-order models that aim to learn more structured and physically interpretable latent representations of fluid flows \citep{solera2024beta}. Second, likelihood-based guidance is less straightforward in the latent setting because sampling is performed in latent space while observations are defined in physical space. This can require repeated decoding and, for more general observation operators, differentiation through the decoder. Even so, if latent representations can be learned that preserve the relevant flow features reliably, latent diffusion may still provide an effective route to scalable SRDA at higher resolutions and in more complex meteorological and geophysical flow applications.

\section{Network architectures}
\label{app:architectures}

\begin{figure}
  \centering
  \includegraphics[width=1.0\textwidth]{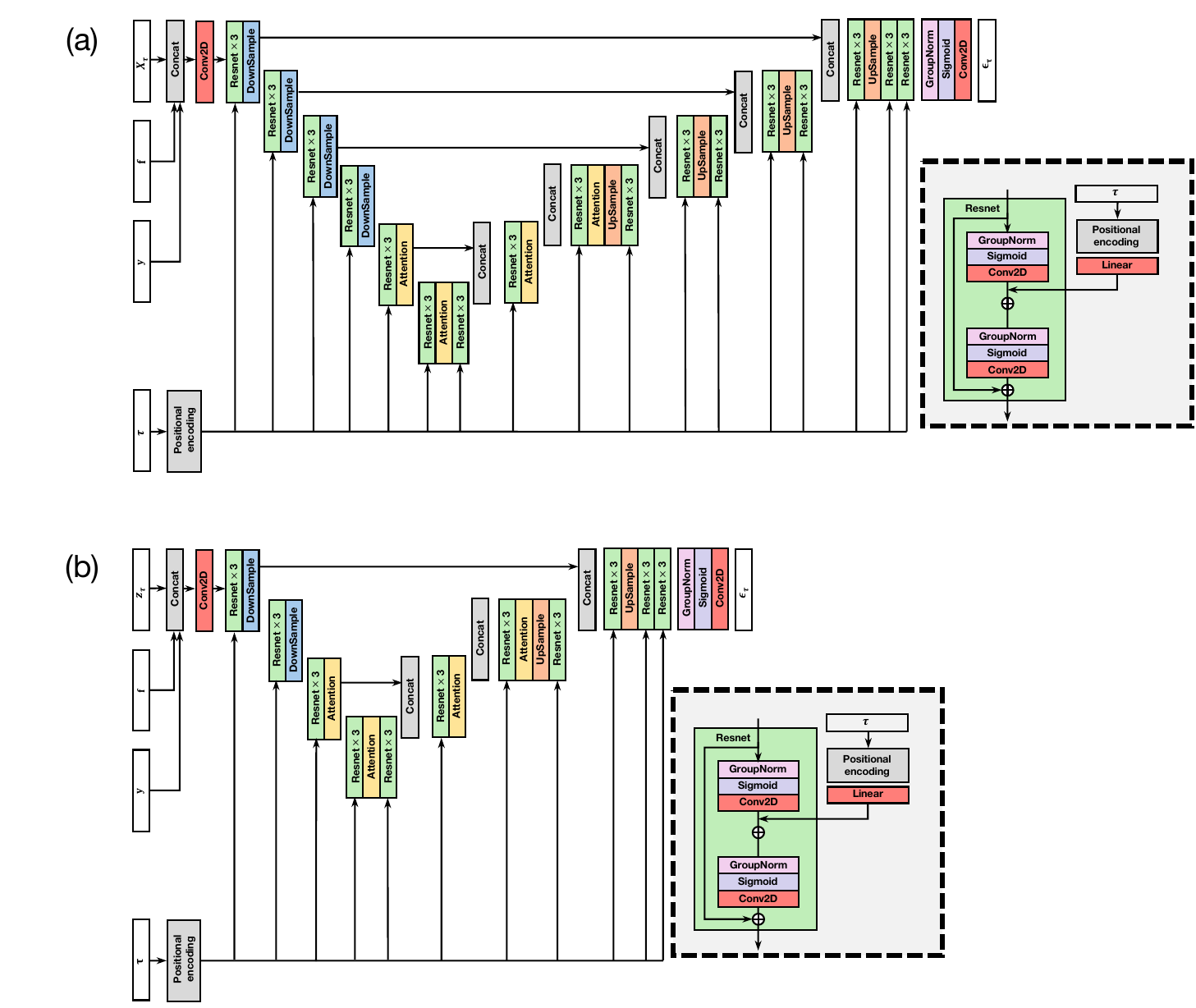}
  \caption{U-Net denoiser architectures used in this study: (a) the pixel-space diffusion model (DiffSRDA) and (b) the latent-space scalability ablation model (LatDiffSRDA).}
\label{fig:model_architectures}
\end{figure}

Figure~\ref{fig:model_architectures} shows the U-Net denoiser backbones used for the pixel-space diffusion model (DiffSRDA) and for the latent-space scalability ablation (LatDiffSRDA). In both cases, the denoiser is trained in the standard $\boldsymbol{\varepsilon}$-prediction form, with diffusion time $\tau$ embedded through a positional encoding and injected into the residual blocks.

For DiffSRDA, shown in Figure~\ref{fig:model_architectures}(a), the noisy diffusion variable is the HR analysis window $\mathbf{X}_\tau$, represented as five channels. The conditioning tensor consists of five additional channels: three channels from LR forecast features, corresponding to three LR times around the assimilation time and upsampled to the HR grid, and two channels from processed observations aligned with the assimilation interval. These conditioning channels are concatenated with $\mathbf{X}_\tau$ along the channel dimension, so the U-Net input has 10 channels in total, and the network outputs the predicted noise field $\boldsymbol{\varepsilon}_\tau$ with five channels, one for each HR frame. The network uses group normalization with 32 groups, and self-attention is applied at the intermediate spatial resolution of 8.

For LatDiffSRDA, shown in Figure~\ref{fig:model_architectures}(b), the SRDA cycle and conditioning structure remain the same, but diffusion is performed in the vector-quantized latent space produced by a VQ-VAE, as described in Appendix~\ref{app:SRDA_ldm}. The noisy diffusion variable is therefore the latent tensor $\mathbf{z}_\tau$ rather than the pixel-space state window. The conditioning channels from the LR forecasts and observations are concatenated with $\mathbf{z}_\tau$, and the U-Net outputs the corresponding predicted latent-space noise field $\boldsymbol{\varepsilon}_\tau$. The encoder maps an HR window to a latent representation, which is quantized using a learned codebook, and the decoder reconstructs the HR window from the quantized latents. The VQ-VAE is trained separately and then kept fixed during diffusion training and sampling. The corresponding VQ-VAE configuration and latent tensor shapes are reported together with the LatDiffSRDA implementation settings.

\end{document}